\documentclass[12pt, preprint]{aastex}
\usepackage{natbib}
\usepackage{graphicx}
\usepackage[usenames]{xcolor}
\usepackage[normalem]{ulem}
\bibpunct{(}{)}{;}{a}{}{,}		

\usepackage{multirow}
\usepackage{threeparttable}

\slugcomment{Not to appear in Nonlearned J., 45.}

\shorttitle{Ionisation  in atmospheres of very low-mass objects}
\shortauthors{Bailey et al.}

\begin{document}


\title{Ionisation in atmospheres of brown dwarfs and extrasolar planets\\
       VI: Properties of large-scale discharge events}

\author{R. L. Bailey\altaffilmark{1,2} , Ch. Helling\altaffilmark{1}, G. Hodos{\'a}n\altaffilmark{1}, C. Bilger\altaffilmark{1, 3}, C. R. Stark\altaffilmark{1} }
\affil{$^1$ SUPA, School of Physics \& Astronomy, University of St Andrews, St Andrews,  KY16 9SS, UK}
\email{ch@leap2010.eu (\today)}
\affil{$^2$ Zentralanstalt f\"ur Meteorologie und Geodynamik, Hohe Warte 38, 1190
Vienna, Austria}
\affil{$^3$ Department of Engineering, University of Cambridge, Cambridge CB2 1PZ, UK}




\begin{abstract}
Mineral clouds in substellar atmospheres play a special role as a
catalyst for a variety of charge processes. If clouds are charged, the
surrounding environment becomes electrically activated, and ensembles
of charged grains are electrically discharging (e.g. by lightning), which
significantly influences the local chemistry creating conditions
similar to those thought responsible for life in early planetary
atmospheres.  We note that such lightning discharges contribute also to
the ionisation state of the atmosphere. We apply scaling laws for electrical discharge processes
from laboratory measurements and numerical experiments to {\sc Drift-Phoenix} model atmosphere results to
model the discharge's propagation downwards (as lightning) and upwards (as sprites) through the atmospheric clouds.  We evaluate the spatial extent and energetics of lightning discharges.
The atmospheric volume affected (e.g.  by
increase of temperature or electron number) is
larger in a brown dwarf atmosphere ($10^8~-~10^{10}$\,m$^3$) than in a
giant gas planet's ($10^4~-~10^{6}$\,m$^3$). Our results suggest that
the total dissipated energy in one event is $<10^{12}$ J for all
models of initial solar metallicity.  First attempts to show the
influence of lightning on the local gas phase indicate an increase of
small carbohydrate molecules like CH and CH$_2$ at the expense of CO
and CH$_4$. Dust forming molecules are destroyed and the cloud
particle properties are frozen-in unless enough time is available for
complete evaporation.  We summarise instruments
potentially suitable to observe lightning on extrasolar objects.
\end{abstract}





\section{Introduction}

Atmospheric electrical discharges are 
observed within our
solar system (e.g. \citealt{desch92,zarka08, dyudina04,
  fischer11}), most notably Earth, where lightning is a common
large-scale discharge phenomenon. There are lesser-known
discharges of even larger scales that occur in Earth's upper
atmosphere such as blue jets and giant red sprites, a fraction of
which is triggered by a lightning strike (e.g. \citealt{bocci95}).

Discharges within the atmospheres of Jupiter and
Saturn have also been detected (e.g. \citealt{rinnert98, little99,
  dyudina04,fischer11}); Jupiter's discharges have been observed and
imaged in the optical and corresponding radio emissions have also been
detected. Saturn is very loud in the radio, giving off bursts of SEDs
(Saturn Electrostatic Discharges) when a discharge (lightning) storm
develops in its atmosphere. Both of these planets have estimated total
discharge energies of $\approx 10^{12}-10^{13}$ J, much larger than
the values of 10$^8-10^9$ J of a single lightning strike
energy on Earth. Observable discharges on these
giant gas planets appear less common and are more localised compared
to Earth.  Electromagnetic signatures from discharges in the
atmospheres of Uranus and Neptune have also been detected
\citep{gurnett90,zarka86}; however, these measurements are sparse and
little is known about the properties of discharges on the outermost
gas planets.

 Electrostatic discharges have long been thought to be a catalyst for
 the creation of prebiotic molecules responsible for the origin of
 life on the young Earth \citep{miller53}, and so the scales of
 discharges and the amount of energy deposited into an exoplanetary
 atmosphere are of great interest. Also of interest is whether the
 discharge events are large and/or strong enough to be detectable from
 afar, as the presence of detectable emissions could reveal
 information on the local physiochemical processes and the chemical
 composition within other atmospheres. In this paper we present the
 first study of the characteristic scales of lightning discharges in
 very cool, low mass objects.

\citet{2012P&SS...74..156Z} estimate that it is not totally
unrealistic to detect lightning on a extrasolar gas giant planet at a
distance of 10pc. It would need to have lightning discharges emitting
$10^5\times$ more energy than Jupiter. Brown dwarfs are of Jupiter's
size and a large fraction are more active than Jupiter due to their
fast rotation which drives atmospheric circulation and cloud formation
processes. While Earth's cloud lightning storms are dispersed across
the planet, Jupiter's discharges are observed to only occur within
certain storm cells. Saturn's SEDs seem to be observable when a single
massive storm (some large enough to contain many Earths) forms within
its atmosphere.

We may argue that indications for lighting, or the electromagnetic
signature of high-energy discharge processes, may have already been
detected in extrasolar substellar atmospheres. While quiescent X-ray
emission decays between objects of spectral class M7 and M9
(\citealt{berg2010}), objects as cool as L5/T7 brown dwarfs exhibit
long-lived H$\alpha$ emission and quiescent radio emission
(\citealt{hall2002,burgasser2011,burgasser2013}). \cite{route2013} and
\cite{williams2013} observed the radio emission from a T6.5 dwarf.
The physical mechanism behind flaring, quiescent X-ray and radio
emission may be the result of the energy release into the ambient
atmosphere associated with reconnecting magnetic field lines.  This
implies a coupling between the bulk, convective motions of the
atmosphere and the ambient magnetic field.  Despite uncertainty
regarding the origin of magnetic fields in fully convective objects,
the more pressing question is which processes contribute to the
ionisation of such ultra-cool atmospheres such that convective energy
can be released by magnetic field coupling. This paper will contribute
to resolving this question by presenting a first study of large-scale
discharge properties in extrasolar, ultra-cool atmospheres.

  Other plasma-initiated emission may also be present in substellar
  atmospheres.  Magnetospheric electrons that are accelerated along
  magnetic field lines will interact with the neutral atmosphere
  stimulating auroral-type emission as suggested in
  \cite{nichols2012}.  However, a seed plasma is also required for
  such auroral emissions which originate from the solar wind, cosmic rays
  and the geologically active Io in the Jupiter system. It might be
  suggestive to think about moons in conjunction with satellite systems
  (that exert tidal forces on the moon) as plasma sources for
  exoplanets, similar sources (incl. host star winds) can not a priori
  be expected to be available to brown dwarfs or brown dwarf
  systems. Hence the same question arises, namely, where does the
  seed plasma come from that drives an aurora.

The onset of lightning is not well understood
(e.g. \citealt{gorm98}). However, streamer discharges are thought to
play a major role and are suggested to determine the early stages of
large-scale discharges like lightning and associated sprites
(\citealt{phelps74,raizer91,gorm98, briels08a}) as they occur in a
variety of ionised media with a large range of pressure and
temperature.

We adopt the idea that a large-scale lightning strike or a large-scale
sprite discharge is composed of various small streamer events, and
that a streamer triggers and develops into such a large-scale
discharge event. This is not always  the case since most of the
gas-discharges in an atmosphere may in fact not even develop into a
streamer (see e.g. \citealt{helling13}). However, in this
paper we are interested in investigating the scales that large-scale
gas-discharges can develop, what atmospheric volume might be affected,
and what amount of energy may be deposited into the atmospheres of
brown dwarfs and planets by large-scale lightning discharges. We
also discuss  how the local gas phase can be affected by the
temperature increase in a discharge channel in the atmosphere.

We utilised scaling laws for discharge processes based on laboratory
measurements \citep{briels08a,briels08b,nijdam08} and numerical
experiments \citep{pancheshnyi05} in order to provide a first
investigation of the spatial extent and energetics of discharges
within the atmospheres of substellar objects, i.e. in brown dwarfs and
extrasolar giant gas planets. In principle, these investigations can
also be applied to smaller planets such as those currently observed by
the {\it Kepler} space mission.

We start with a summary (Sect.~\ref{s:clouds}) of our work on clouds
in the atmospheres of brown dwarfs and giant gas
planets. Section~\ref{sec:method} introduces the scaling laws which we
use and outlines our method of applying these scaling laws to brown
dwarf and giant gas planet atmospheres. Section~\ref{s:results}
summarises our results. Lightning on Earth produces a large
  number of observable signatures across the energy spectrum. We
  summarise these signatures in Sect.~\ref{sec:propsobs} and collate
  possible instruments for their detection on Brown Dwarfs or
  exoplanets.

\section{Cloudy Substellar Atmospheres}\label{s:clouds}

Atmospheres of very cool, substellar objects like brown dwarfs (BD)
and giant gas planets (GP) are cold and dense enough that cloud
particles can condense from the atmospheric gas.  The formation of
dust by seed formation and bulk growth takes place in a temperature
window of $\approx~500-2100$~K and leads to the formation of mineral
clouds.  Gravitational settling, convective mixing and element
depletion are major processes that occur in such atmospheres. Helling,
Woitke \& Thi (2008) have shown that the upper cloud region
(low-temperature and low-pressure) will be dominated by small, dirty
(i.e. inclusions of other materials) silicate grains with inclusions
of iron and metal oxides; and the warmer, denser cloud base by bigger,
dirty iron grains with metal inclusions. The actual size of the cloud
particles deviate from this mean value according to a height dependent
size distribution (Fig. 8 in Helling, Woitke \& Thi 2008). The
chemical material compositions as well as the local grain size
distribution change with height inside a cloud in a quasi-stationary
environment. These cloud particles can be charged \citep{helling11b}
and the resulting electric field may be sufficiently strong to
initiate small-scale streamers that develop into large-scale discharge
processes like lightning (\citet{helling11a, helling13}).

 We utilise one-dimensional atmosphere models in what follows, hence,
 we do not take into account any horizontal motions that are so
 obvious on Jupiter and are stipulated for irradiated, extrasolar
 planets from works by Showman et
 al. (e.g. \citealt{Showman2013}). Such horizontal motions will
 produce patchy cloud coverage rather than a homogeneous cloud
 coverage as assumed in 1D models. More complicated atmospheric
 structures involving winds and dynamic meteorology will introduce
 additional effects such as ionization via Alfv\'{e}n ionization
 (Stark et al. 2013).  

\subsection{Atmospheres and Clouds}

We use {\sc Drift-Phoenix} model atmosphere structures
\citep{dehn07,helling08a,helling08b,witte09,witte11} as input for the
local gas temperature and gas pressure.  {\sc Drift-Phoenix} model
atmospheres are determined by a coupled system of equations describing
radiative transfer, convective energy transport (modelled by mixing
length theory), chemical equilibrium (modelled by laws of mass
action), hydrostatic equilibrium and dust cloud formation. The dust
cloud formation model includes a model for seed formation
(nucleation), surface growth, evaporation of mixed materials and
gravitational settling (Woitke \& Helling 2003, 2004; Helling \&
Woitke 2006; Helling, Woitke \& Thi 2008; Helling \& Fomins 2013). The
results of the {\sc Drift-Phoenix} model atmosphere simulations
include the gas temperature - gas pressure structure ($T_{\rm gas}$,
$p_{\rm gas}$), the local gas-phase composition, the local electron
number density ($n_{\rm e}$), the number of dust grains ($n_{\rm d}$)
and their mean sizes ($<a>$); which are all dependent on atmospheric
height. These models are determined by the effective temperature,
$T_{\rm eff}$, the surface gravity, $\log(g)$, and the initial element
abundances which are set to solar values unless specified otherwise.

Atmospheres of varying parameters were used: The effective
temperatures, which represent the total wavelength-integrated
radiative flux, ranged from 1500 K to 2000 K. Within each effective
temperature category, the substellar atmospheres split into categories
of brown dwarfs (BDs; $\log(g) > 4.0$) and giant gas planets (GPs,
$\log (g) < 4.0$), for which we consider solar metallicity ([M/H]$ =
0.0$) and sub-solar metallicity ([M/H]$ = -3.0$) models. The initial
element abundances are oxygen-rich, i.e. more oxygen than carbon is
available.

Dusty cloud layers are expected to form within these substellar
atmospheres. With the formation of initial seed particles, chemicals
can gather on their surface and dust grains grow while
simultaneously depleting the local elemental abundances. As the grains
grow in mass, they fall down through the cloud
and gravitationally settle in the lower layers. Convection in the atmosphere allows
for the constant replenishment of chemicals in the cloud for the
growth of dust grains, as well as the mixing of dust grains of
different size and composition \citep{helling08c}.

\begin{figure}[htbp]
\centering
\includegraphics[scale=0.8]{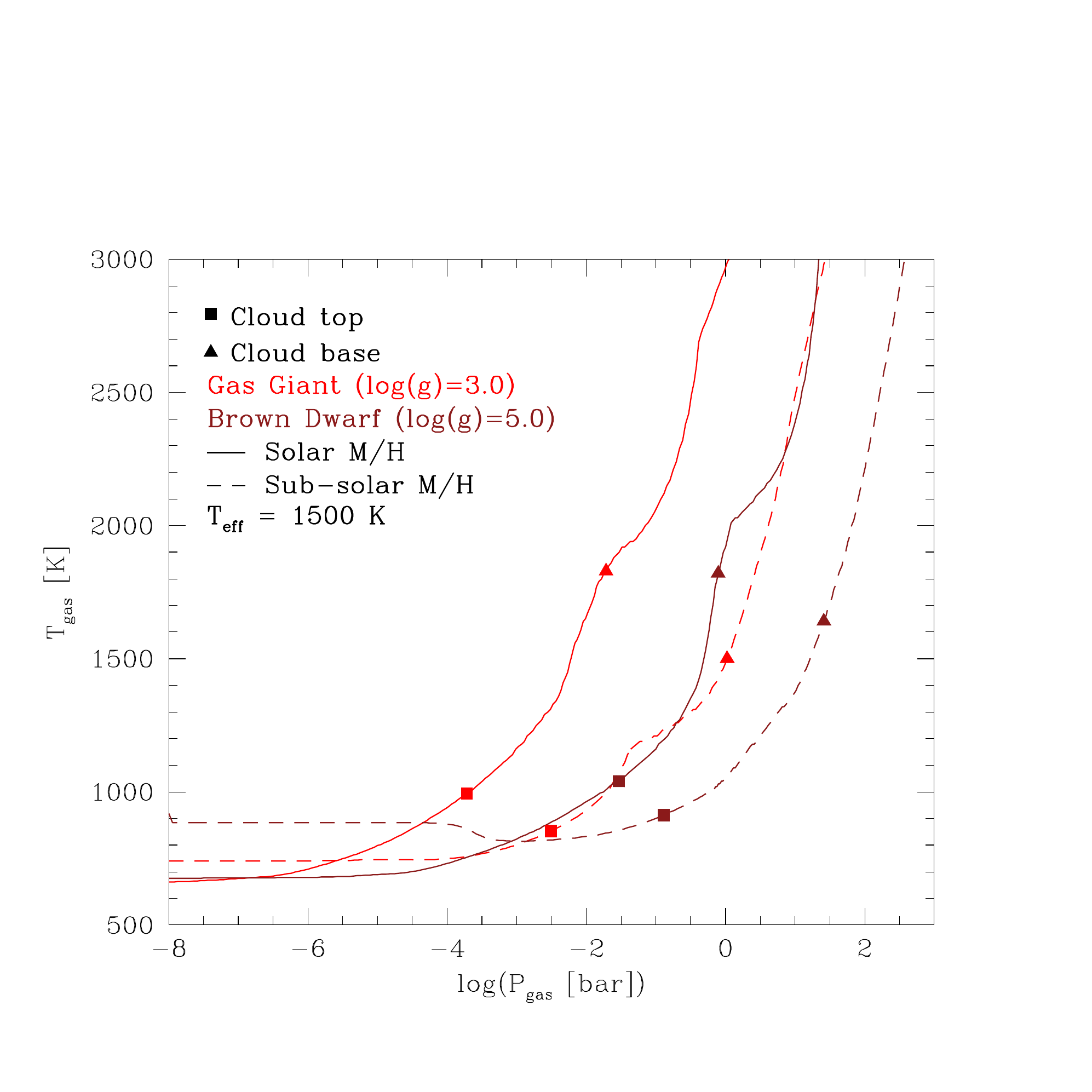}\\*[-2cm]
\caption{Cloud layer boundaries (square = cloud top,
  triangle = cloud base) in  {\sc Drift-Phoenix} model atmosphere structure of brown dwarfs (brown) and giant gas planets (red) with $T_{\rm eff}=1500$ K. Over-plotted are two cases of different metallicity (solar, [M/H]$ =
  0.0$ -- solid lines; sub-solar, [M/H]$ = -3.0$ -- dashed lines).}
\label{fig:pTplot}
\end{figure}

\begin{figure}[htbp]
\centering
\includegraphics[scale=0.8]{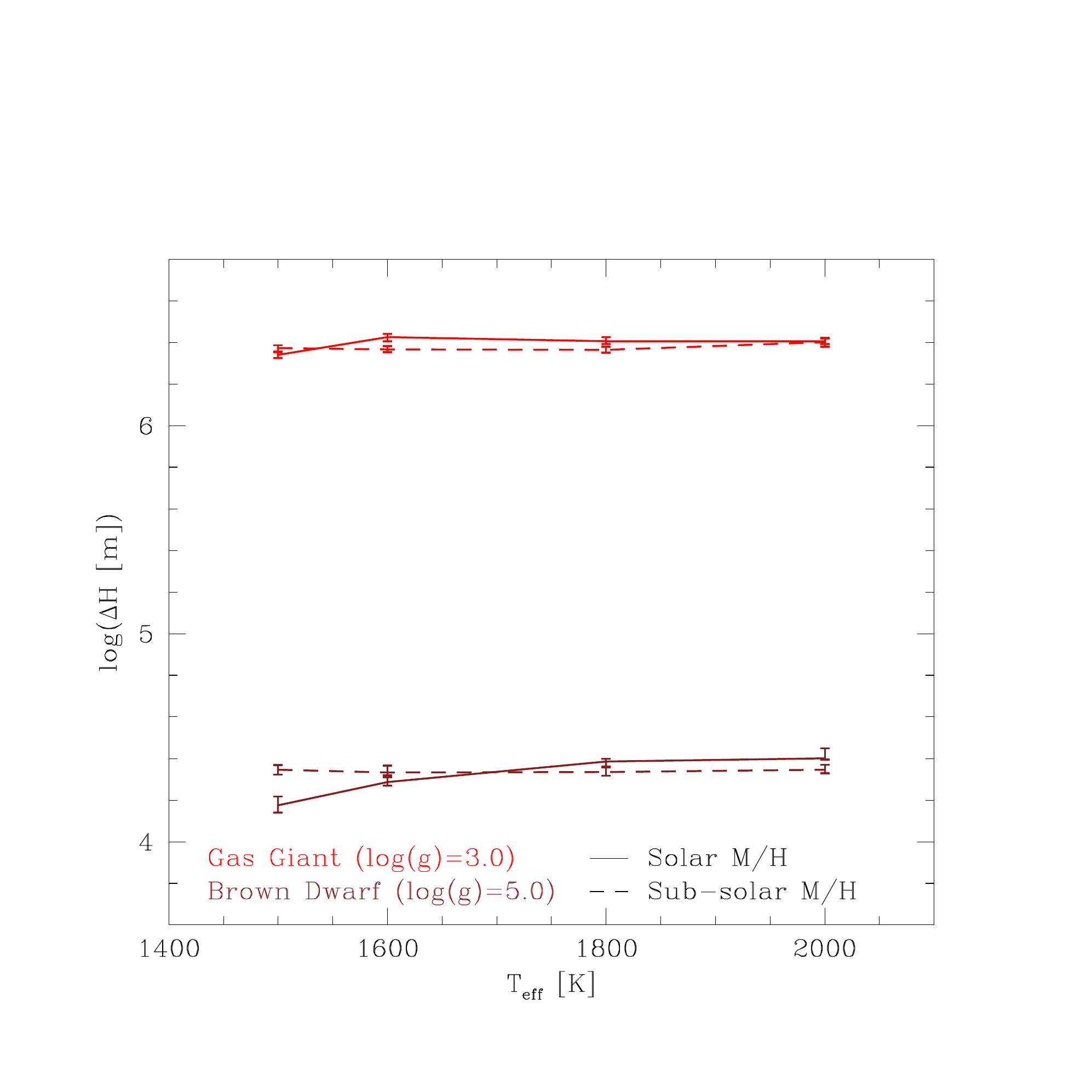}\\*[-2cm]
\caption{Vertical cloud extension, $\Delta H(T_{\rm eff})$, for {\sc Drift-Phoenix} model
  atmospheres of different $T_{\rm eff}$ with $ \log (g)$ = 3.0 (GP,
  red), 5.0 (BD, brown); [M/H] = 0.0 (solid lines), -3.0 (dashed
  lines). The error bars indicate the uncertainty with which the cloud
  height is determined based on the {\sc Drift-Phoenix} atmosphere
  models (see also Sect~\ref{ss:errors}).}
\label{fig:deltaH}
\end{figure}

Due to the 1D nature of the models, the clouds are assumed to form in
horizontally extended layers within the atmospheres. The cloud deck is
here defined as the first particle nucleation maximum and the point at
which all cloud particles have evaporated due to the locally high
temperature\footnote{See Eq. 16 in \cite{woi2004} for a different
  definition of the cloud height.}. These cloud decks are usually of
the order of 10$^7$ m in vertical extension ($\Delta H$) in the GPs
and 10$^4$ m in the BDs due to the higher surface gravity
(Fig. \ref{fig:deltaH}). Clouds in low-metallicity atmospheres form at
lower temperatures and higher pressures than clouds in the
solar-metallicity atmospheres (\citealt{witte09}). GP atmospheres also
form clouds at much lower pressures than their BD counterparts
(Fig. \ref{fig:pTplot}). With increasing effective temperature, the
clouds also form at decreasing pressure, hence they are located at
higher atmospheric altitudes.

\subsection{Large-scale charge separation within clouds}

Charge separation requires motion and friction for the separation and
relocation of charges to occur. This can be provided by convective and
turbulent motions, which are common in atmospheres and cloud regions,
and have been observed in the cloud and storm systems of Jupiter and
Saturn.  Large-scale motions (global circulation patterns) on
exoplanets are inferred from thermal emission observations from
infrared lightcurves e.g. \citep{knu2012} and from simulations
e.g. \citep{heng2011, dd2012, rauscher2012, perna2012, show2013}.  All
suggest a displacement of the hot-spot (a global temperature maximum
of the atmosphere due to irradiation by the host star) as result of a
fast eastward windflow at the equator that displaces the thermal
maxima to the east. Resulting differential, height-dependent rotation
of the atmosphere can naturally be expected. Collisions between the
cloud particles can lead to tribo-electric charging, which occurs due
to friction as one material surface rubs against another and charge is
transported from one grain to another.  Experiments show that in
systems of colliding particles, negative charge moves to the smaller
particles while the larger particles become positively charged (Lacks
and Levandovsky, 2007; Krauss et al., 2003; Zheng et al.,
2003). Larger particles will sink faster to the bottom of the cloud
while smaller particles remain longer at the top and can be easily
transported by winds, establishing large-scale charge separation
within the cloud. Similar scenarios are suggested in
\citet{merrison2012, zarka04}, and in \citet{farrell99}.

Another mechanism at work in very turbulent atmospheres would be
fracto-emission. Fracto-emission is the emission of particles and
electrons during and after the fracture of a dust grain due to
external stresses.  This process results in the emitted, fractured
material acquiring different charges~\citep{dickinson84}. The
different transport properties of the charged grains cause them to
migrate within the dust cloud and an electric field can be
established. The shattering process depends strongly on the relative
velocities involved in the collisions.

\section{Application of gas-discharge scaling laws\\ to brown dwarf and giant gas planetary atmospheres}
\label{sec:method}

We aim to investigate the geometrical extension of potential discharge
processes in BD and GP atmospheres, and to find out where potential
lightning discharges occur.  We utilise experimentally obtained
scaling laws ({\it similarity relations}) describing discharge properties to provide a first
insight into the potential scale size of discharge processes in BDs
and GPs and derive values for the energy deposited and the atmospheric
volume affected. Knowing both the pressure-temperature scales in our
model atmospheres and the cloud extensions (from {\sc Drift-Phoenix}
atmosphere simulations) enables us to estimate how discharge
properties (such as the discharge propagation length, the radius of
the discharge channel or the gas volume affected by the discharge)
scale within the modelled cloud layers.

We first outline our modelling ansatz in
Sects.~\ref{ss:modelansatz}--~\ref{ss:localfield}. Section~\ref{sec:scalinglaws}
introduces the scaling laws which we utilise.


\subsection{Modelling ansatz}\label{ss:modelansatz}
To investigate atmospheric discharge events we model the top and
bottom of the cloud layer as two equal and oppositely charged surfaces
analogous to the parallel plates of a capacitor.
This ansatz has been applied by \cite{raizer98}, \cite{yair09} and by
\cite{pasko00} to study sprites in solar system planetary atmospheres
and fractal streamer propagation. It serves as a
first-order-approximation to investigate the latent physics and
chemistry of the system. It is assumed that a streamer-initiated
discharge can occur between two charge-carrying surfaces (i.e. the
cloud top and the cloud base) if the build-up of electric charge is
large enough for the resulting electric field to overcome the local
breakdown field.

The net charge on the cloud top and base (and hence the resulting
electric field) is unknown, unless all necessary charging processes
can be modelled. Therefore, a two-fold strategy was followed to
  evaluate the electric field which was inspired by previous works
  (\citet{raizer98,pasko00,yair09}).

In case (i) $Q=$const: the local electric field is
evaluated by assuming a constant  charge $Q$. The discharge
would then propagate from a point in the cloud, $z_{\rm init}$ at
which the local electric field $E(z_{\rm init}, Q)$ corresponding to charge $Q$, 
exceeds the break-down field strength $E_{\rm th}$: $E(z, Q) > E_{\rm th}$. The exact position
of this point and the distance from the cloud top varies between the
different model atmospheres. 

In case (ii) $z_0=$ const: the discharge is initiated
for the charge $Q$ that fulfils the break-down criterion $E(z_0, Q) > E_{\rm th}$
at a fixed point, $z_0$,  below the cloud top. This results in the minimum
amount of charge, $Q_{\rm min}$, required to initiate a discharge at a
fixed point. From this initiation point the discharge propagates into
a rising breakdown field. However, due to the enhanced field at the
streamer tip it can continue  to propagate; therefore, from the point of
initiation the propagation is independent of the initial ambient field.

\begin{figure}[htbp]
\centering
\includegraphics[scale=0.8]{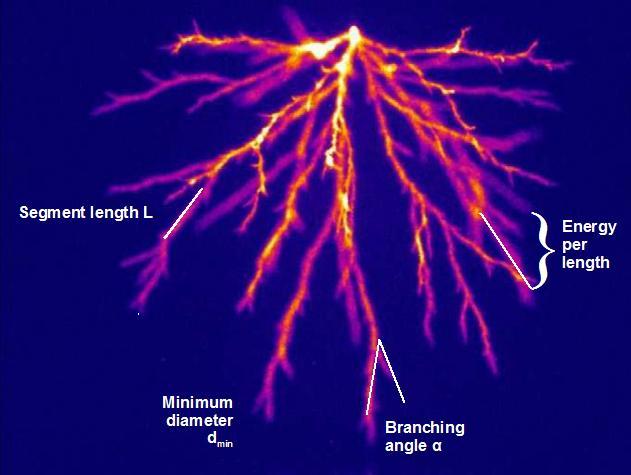}
\caption{This image is adopted from  \citet{briels08b} [{\small \copyright IOP Publishing.  Reproduced with
permission.  All rights reserved.}] to visualise the streamer
  properties evaluated in laboratory experiments and used
  in this paper: The segment length, $L$ (Eq.~\ref{eq:length}), is the
  length of a single segment of the streamer. The
  minimum diameter, $d_{\rm min}$ (Eq.~\ref{eq:dminT}), is the minimal
  segment diameter seen on a streamer. The branching angle, $\alpha$
  (Eq.~\ref{eq:branching}), is the angle between two branches from the
  same parent segment. The energy per length, $E_{\rm tot}/l$
  (Eq.~\ref{eq:ELrelation}), is the amount of dissipated energy per
  length of single segment.}
\label{fig:properties}
\end{figure}

In both cases, (i) and (ii), we assume horizontal homogeneity of the
ambient gas; therefore, only the streamer scaling into the vertical
direction is considered. This assumption is reasonable since
on Earth electric currents flow in the atmospheric electric field
preferentially upwards, towards the apex of the geomagnetic field line
(\citet{ryha2012}). The propagating discharge has the form of a
sprite-like discharge, starting with only a few branches that
split into more and more filaments
(Figs.~\ref{fig:properties}).

The discharge propagates through the atmospheric gas according to
scaling laws (Sect.~\ref{sec:scalinglaws}). Each discharge starts with
a characteristic diameter and as the discharge evolves and branches,
it does so into branches of progressively smaller diameters until the
minimum diameter is reached. At this point the whole discharge event
stops. This approach follows \citet{ebert10} and their description of
streamer behaviour.

\subsection{Breakdown field}\label{ss:Eth}

The breakdown field, $E_{\rm th}$, is the minimum threshold electric
field that must be overcome in a medium for electrical breakdown to
occur. For electric fields above this value, the gas ionisation rate
exceeds the electron attachment rate and the ionisation front can
propagate. The breakdown field changes depending on the composition of
the surrounding medium and the product of the gas pressure and the
distance between the electrodes. We use the same description for the
breakdown field as in \citet{helling13},
\begin{equation} \label{eq:ebr}
\frac{E_{\rm th}}{p} = \frac{B}{C+\ln(pd)},
\end{equation}
where $B$, $C$ and $pd$ are constants and values are summarised in
\citet{helling13}.  Eq.~\ref{eq:ebr} defines the
  breakdown criterion as a function of gas pressure $p$ and electrode
  separation $d$.  For a given gas composition there exits a critical
  value of $(pd)_{\rm min}$ that yields the minimum electric field
  strength for electrical breakdown $E_{\rm th,min}$.  For values of
  $pd<(pd_{\rm min})$ the breakdown field $E_{\rm th}$ decreases with
  increasing $pd$; and for values of $pd>(pd)_{\rm min}$, the
  breakdown field $E_{\rm th}$ increases with increasing $pd$.  Here,
we use the parameters initially determined for Jupiter's atmosphere
(Sect. \ref{sec:properties}). We evaluate the dependence on different
chemical composition of the atmosphere in Sect. \ref{sec:diffatmo}.

\begin{figure}
\centering
\includegraphics[scale=0.8]{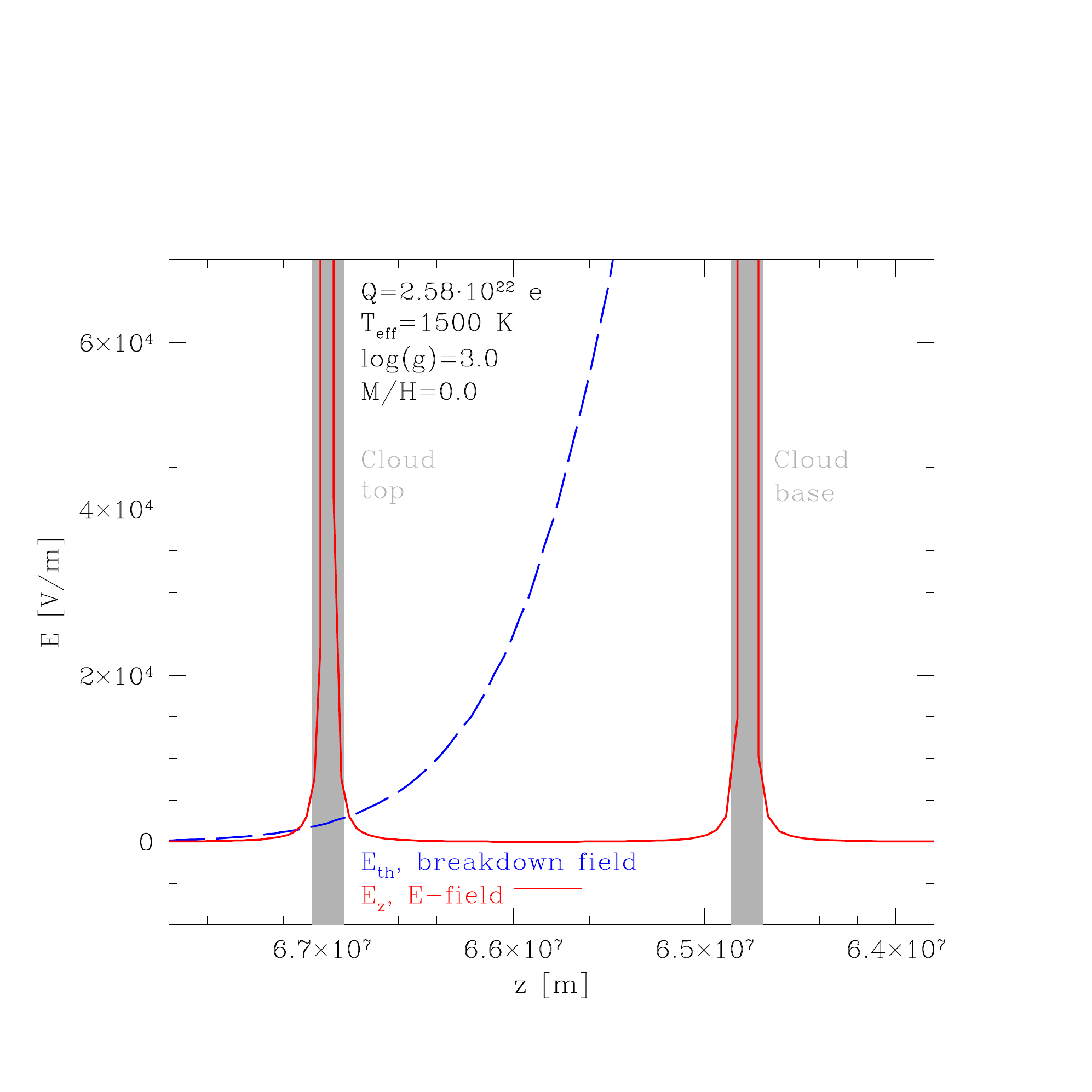}\\*[-1cm]
\caption{The local electric field (red) and the breakdown field (blue)
  between the charged top and bottom of the cloud layer (depicted in
  grey) for a {\sc Drift-Phoenix} giant gas planet atmosphere model
  with $T_{\rm eff}=1500$ K, $\log (g) =3.0$ and [M/H]=0.0. Top and
  bottom of the cloud are assumed to carry a charge of $2.58 \times
  10^{22}$e of opposite polarity. This is the minimum amount of charge
  required for a discharge to occur (case (ii)). The discharge occurs at
  the height $z$ at which $E(z) > E_{\rm th}$.}
\label{fig:cloudEplot}
\end{figure}

\begin{figure}
\centering
\includegraphics[scale=0.8]{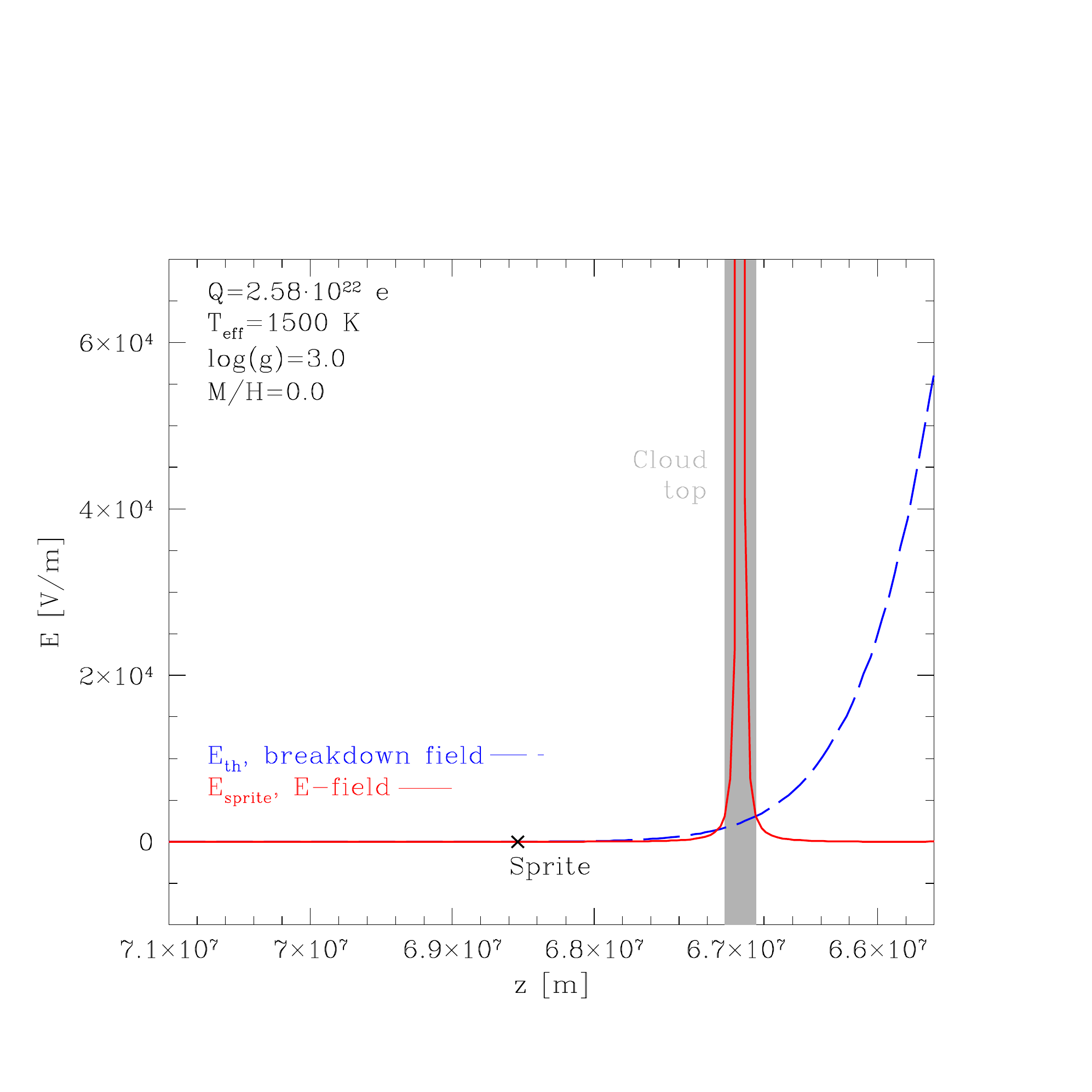}\\*[-1cm]
\caption{The electric field and the breakdown field above the cloud,
  which is an enlarged portion of the field shown in
  Fig. \ref{fig:cloudEplot}. At a distance of $\sim 10^6$ m above the
  cloud, the electric field exceeds the breakdown field suggesting
  that a sprite may occur here. 
}
\label{fig:SpriteE}
\end{figure}

\subsection{Model electric field for atmospheric discharges}\label{ss:localfield}
For simplicity we model the electric field of the cloud with net charge $Q$ as a simple electric dipole,
\begin{equation} \label{eq:Ez}
   E(z) = \frac{Q}{4 \pi \epsilon_0} \left [ \frac{1}{(z -
       z_{base})^2} - \frac{1}{(z - z_{top})^2} \right ].
\end{equation}
A charge distribution within a cloud that has undergone large-scale
charge separation can be complex, with opposing layers of charge not
always sitting parallel to the horizontal axis but also beside each
other due to potentially complex convective cells. Multiple pockets of
localised charge distributed across the cloud could also be a likely
situation \citep{rakov03}. For simplicity and ease of comparison with
similar studies \citep{raizer98,yair09}, the dipole electric field in
Eq. \ref{eq:Ez} of two poles (or small pockets of opposite charge) was
adopted to emulate a simplified Earth storm cloud cell as suggested by
\citet{rakov03}. The electric dipole field should provide a fair approximation due
to the potentially large distances between the charge centres. Both the local
field and the breakdown field are shown in Fig. \ref{fig:cloudEplot}
(for case (ii)): $Q=Q_{\rm min}$), and a discharge starts at the point in the
atmosphere where $E(z) > E_{\rm th}$.

The same model is used to study if and where sprites may occur above
the cloud at where $E(z) > E_{\rm th}$
(Fig. \ref{fig:SpriteE}). Current sprite theory suggests that these
forms of electrostatic discharge occur when charge below is removed
suddenly by a lightning stroke, so that a quasistatic field appears
above the cloud where a single charge centre remains
(e.g. \citet{gorm98, briels08b}). It is expected that a mirror charge
may appear in a conducting ionosphere (see \citet{raizer98}); however,
no ionospheric considerations have been included here due to lack of
knowledge on the nature of ionospheres in brown dwarfs and extrasolar
giant gas planets. Sprites occur only milliseconds after powerful
lightning strikes within this quasistatic electric field above the
cloud.

Although both the local field and the sprite field decay above the
cloud, there is a point at which the {electrical breakdown condition
  is satisfied (see Fig.~\ref{fig:SpriteE}). We assume that a sprite launches at this point,
  triggered by a lightning discharge at lower altitudes.

\subsection{Scaling laws for Streamers and Sprites} 
\label{sec:scalinglaws}

Various laboratory experiments in recent years
\citep{briels08a,briels08b,nijdam08,pancheshnyi05} 
studied the properties of the basic discharge instability which occurs
in the form of streamers.  Streamers are electron avalanches,
initiated by fast-moving free `seed' electrons that have been
accelerated by a sufficiently strong external electric field. The
initial electrons acquire enough energy to knock further electrons
from the ambient atom or molecules, which liberate more electrons and
the event cascades into an avalanche. This evolves into a
self-propagating ionisation front that advances through the medium
evolving into a discharge.  Laboratory experimental studies of
streamers have identified empirical scaling laws that relate
characteristic properties of their evolution.  This enables us to
apply such scaling laws to physically similar systems such as those
found in BD and GP atmospheres. Similar investigations have been made
in the upper atmosphere of Earth, where the properties of sprites have
been quantified \citep{pasko97, gerken00, cummer06,
  stenbaeknielsen07}. Sprites, which are massive discharges that occur
above thunderstorms milliseconds after powerful lightning strikes,
have a similar filamentary structure to streamers. It has therefore
been suggested \citep{briels06,ebert10} that streamers and sprites
share similar mechanisms inferring that sprites are streamers scaled
to atmospheric pressures.

Empirical scaling laws for the following streamer properties were
experimentally determined: the streamer segment lengths ($L$), the
minimum diameters ($d_{\rm min}$), the total energy ($E_{\rm tot}$)
and volume ($V_{\rm tot}$) of a discharge event
(Fig.~\ref{fig:properties},
\citet{briels08a,briels08b,nijdam08}). These were found to scale with
the gas pressure and applied voltage.

\noindent
\underline{\emph{Electron mean free path:}} The streamer length was
observed to scale with the local gas pressure. The physical reason is
that a streamer is a flux of electrons travelling through an ambient
gas: the mean free path of a single electron before it hits a neutral
gas particle is:
\begin{equation} \label{eq:mfp}
l_{\rm mfp} = (\sigma n)^{-1},
\end{equation}
and hence, $l_{\rm mfp} \propto n^{-1}$ or $l_{\rm mfp} \propto p_{\rm
  gas}^{-1}$. $\sigma$ is the collisional cross sectional area for an
electron-neutral interaction and $n$ is the number density of the gas,
which is related to the gas pressure $p_{\rm gas}$ by the ideal gas
law $p_{\rm gas}V = nk_{B}T_{\rm gas}$.  This relates to the Paschen
curves, which plot the breakdown voltage of a gas as a function of
$pd$, the product of the gas pressure and separation of the capacitor
electrodes (for a summary see Helling et al. 2013).  For values of
$pd<(pd)_{\rm min}$ the breakdown voltage decreases with increasing
$pd$; and for values of $pd>(pd)_{\rm min}$, the breakdown voltage
increases with increasing $pd$.

 At high pressures, the mean free path is smaller with respect to the
 electrode separation resulting in more collisions during the
 electrons transit between the electrodes. Each collision randomises
 the electrons motion and will reduce the electrons energy. This means
 that the electrons energy may be insufficient to ionise the neutrals
 it collides with; therefore, requiring a larger voltage to insure
 sufficient electron energization for electrical breakdown to occur.

At low pressures, the electron mean free path is larger with respect
to the electrode separation and the electrons will participate in
fewer collisions. In this scenario, the electron may retain its energy
but will have fewer collisions requiring a greater breakdown voltage
to insure that the collisions that occur are ionising.

The lowest breakdown voltage is found at the value of $pd$ where these
two competing effects balance. While the breakdown voltage depends on
the gas pressure, it also depends on the type of gas as each species
has a different ionisation energy.

\noindent
\underline{\emph{Minimum diameter}:} The minimum diameter is the
minimal width of a streamer segment, below which it does not propagate
(Fig. \ref{fig:properties}). Minimal diameter streamers do not branch
into further segments, and are observed at the very final tips of
streamers. The assumption that streamers have a minimal diameter is
explained in \cite{briels08b}: the streamer tip consists of a space
charge layer, which causes an enhancement of the local electric field. The
size of the space charge layer is defined by the inverse of the
maximum of the Townsend ionisation coefficient, which is derived from
the molecular properties of the gas and changes with pressure. The
diameter needs to be larger than this space charge layer to propagate
farther, hence there is a minimum possible streamer diameter  for
a given gas and gas pressure. Most empirically derived relations are of
the following form,
\begin{equation} \label{eq:dmin}
pd_{min} = A \; [mm~bar],
\end{equation}
The value for the constant $A$ varies between authors. This relation
was tested in detail by \cite{briels08b}, where they found results of
$A_{air} = 0.20 \pm 0.02$ to $0.30 \pm 0.02$ in air and $A_{N_{2}} =
0.12 \pm 0.03$ in an N$_{2}$ gas. \cite{dubrovin10} expanded on this
by looking at the properties of planetary gas mixtures. For a Jovian
mixture they found $A_J = 0.26 \pm 0.03$, for a Venusian
mixture $A_V = 0.09 \pm 0.03$ and a value of $A_{air} = 0.12 \pm 0.03$
in air, similar to \cite{briels08b}.

\cite{briels08b} studied images of sprites in the upper terrestrial
atmosphere and evaluated a height-dependent minimum diameter. From
this, the dependence of the local gas temperature obeys the relation
\begin{equation} \label{eq:dminT}
   \frac{p d_{min}}{T} = A \left [\frac{mm~bar}{293K} \right] .
\end{equation}
The temperature dependence in Eq.~\ref{eq:dminT} is predicated on the
product $nd$, and can be written as $nd = p d/(k_B T)$ assuming an
ideal gas. We note that the ideal gas law can only be applied to the
ambient gas into which the streamer travels, not to the gas that is
affected by the streamer.  At standard temperature and pressure,
Eq. \ref{eq:dminT} applies to streamers and is consistent with the
experimental values in \cite{briels08b}, supporting the assumption
that sprites and streamers are similar in nature.  

 All experimental works cited below have been performed under
  normal pressure on the Earth surface ($\sim 0.1\,\ldots\,1$bar) and
  for gases of a nitrogen/oxygen mixture or pure
  nitrogen. \cite{helling13} demonstrated that different gas
  composition do not significantly influence the electric field
  breakdown in the astrophysical systems studied here. This leads us to assume that the scaling laws
  applied  will not significantly be affected by the different
  gas-phase composition in our extrasolar atmospheres.

\noindent
\underline{\emph{Segment Length}:} The length $L$ is the value of
the diameter-dependent segment length in air as suggested in
\citet{briels08b}. It describes the distance a single segment travels
after a branching event and before the segment itself branches.

\begin{equation}\label{eq:length}
\frac{L}{d} = 11 \pm 4,
\end{equation}
with $d$ being the segment diameter. The values for the segment
length, $L$, change slightly depending on the gas mixture, and a value
of $9 \pm 3$ was found for streamers in an N$_2$ gas. This relation is
pressure independent.  The error estimates given above result from
  experimental error estimates given by the  referenced authors.

\noindent
\underline{\emph{Branching angle}:} The branching angle, $\alpha$, is
the angle between new segment branches when a single segment breaks
into two new segments. It was investigated by \citet{nijdam08} using
3D images of laboratory streamers and the approximately Gaussian
  distribution with a mean angle was found to be,
\begin{equation}\label{eq:branching}
\alpha = 43.0 \pm 12.3^{\circ}.
\end{equation}
 The error estimate results from experimental estimates given in
  \citet{nijdam08}.  We note that streamer channels can also reconnect
  due to different polarities of the channels (\cite{Ebert2010}) which
  we can not take into account in the model
  presented here. \cite{mcharg2010} observe sprite events above
  mesoscale thunderstorms and show that propagating streamer heads are
  both smaller in width and dimmer than splitting streamer heads. The
  reason for streamer head splitting is the development of a Laplacian
  instability caused by an increasing electric field in the streamer
  head (for more details see \cite{mcharg2010}).

\noindent
\underline{\emph{Energy}:} The total energy, $E_{\rm tot}$, of a whole
discharge event
is calculated by looking at each of the individual segments and their
lengths. The value for the total dissipated energy per length is
taken to be (\cite{krider1968, cooray1997, rakov03}):
\begin{equation}\label{eq:ELrelation}
\frac{E_{\rm tot}}{l} = 10^5 \mbox{J\,m$^{-1}$},
\end{equation}
with $l$ a unit length in [m].  The exact number in
  Eq.~\ref{eq:ELrelation} depends on the details of the lightning
  process, e.g. if the first return stroke channel is considered
  (\citealt{cooray1997}). The energy will be dissipated into the heating
  of the discharge channel and the ambient gas around the channel
  (e.g. \citealt{borovsky1995}). \cite{paxton1986} suggest that 70\% of
  the total energy input into a channel is optical radiation from
  the channel. \cite{MacLachlan20013}
  demonstrate that the energy transfer calculation in a discharge
  channel needs to take into account a whole variety of collisional
  processes as for example elastic and inelastic scattering,
  metastable excitation, ionisation, metastable ionisation,
  electron-ion recombination. Hence, the precise value in
  Eq.~\ref{eq:ELrelation} may differ between authors.

   \cite{briels08b} observe that the streamer intensity increases with
  further branching and increasing segment diameter. In comparison,
  the total energy dissipated per event is estimated to be
  $10^7-10^9$J on Earth and $\approx 10^{12}$ J on Jupiter and Saturn.

\noindent
\underline{\emph{Initial diameter}:} The behaviour of streamers
scaling with voltage was described in \citet{briels08a,briels08b},
where it was observed that higher voltages led to more intense, longer
streamers with thicker branches. The work also showed that the segment
diameter increased with increasing voltage, i.e. $d~\propto~V$. However,
no empirical relation was derived. Following this result, we assume a voltage dependence to estimate the  initial diameter for
streamer propagation in our calculations
\begin{equation} \label{eq:dinit}
d_{\rm init} = n_{V} V_{\rm init},
\end{equation}
where n$_{V}=10^{-8}$ is a constant  and is an estimate of the diameter-voltage relation taken from a linear fit to Fig. 5 in \cite{briels08b}. 
This expression is of practical interest to us since $d_{\rm init}$ determines the final total length of the discharge.

\subsection{Model for large-scale discharge structures}\label{ss:lsmodel}

We adopt the idea that a large-scale lightning strike or a large-scale
sprite discharge is composed of multiple streamer events that evolve
into such  large-scale discharge phenomena.   This may not always to be the case and most of the
gas-discharges in an atmosphere may in fact not even develop into a
streamer as discussed in \cite{helling13}. However, in this
paper we are interested in  the scale sizes that large-scale
gas-discharge events can occur for  in substellar atmospheres 
assuming  that they occur.

The empirical scaling laws for streamers (Sect. \ref{sec:scalinglaws}) will
be evaluated for the given set of model atmospheres in
Sect.~\ref{s:results}. After the electric field at a fixed point below
the charged cloud top is calculated (Eq.~\ref{eq:Ez}), the initial
diameter of the streamer, $d_{\rm init}$ (Eq.~\ref{eq:dinit}), is
evaluated.  From this initial diameter the streamer branches after
each segment length, $L$ (Eq.~\ref{eq:length}), into two new segments
with a separation angle, $\alpha$ (Eq.~\ref{eq:branching}).  This approach follows
\cite{briels08a}, where bright streamers were described as starting
with thick diameters that continued to branch into thinner streamers
with smaller diameters until the minimum diameter was reached, at
which point propagation stops. The diameters of each new
segment follow an area conservation law,
\begin{equation}
d_{\rm new} = \sqrt{1/2} \frac{d_{\rm old}}{d_{\rm min,old}} d_{\rm min,new}.
\end{equation}
which depends on the local pressure by Eq.~\ref{eq:dminT}. This ensures
that 
lengths are smaller when moving into higher pressures and
increase if moving into lower pressures. 

To find the minimum diameter of each segment given in
Eq. \ref{eq:dminT}, the parameter $A$  is taken from \citet{dubrovin10} for a Jovian
atmosphere at room temperature, which we assume to be most
similar (hydrogen-based, initially oxygen-rich) to an extrasolar atmosphere,
\begin{equation}
\frac{p d_{\rm min}}{T} = 0.26 \pm 0.03 \left [\frac{mm~bar}{293 K} \right ].
\end{equation}



The total length of the discharge, $L_{\rm discharge}$ (distance
between the initiation point and termination point of the discharge),
and the width of the discharge, $ 2\,W_{\rm discharge}$, were
evaluated using the segment length given in Eq.~\ref{eq:length} and
the branching angle in Eq.~\ref{eq:branching}. We assume that the two
new branches always split at equal angles of $\alpha/2$ relative to
the vertical axis; therefore,
\begin{eqnarray} \label{eq:totallength}
L_{\rm discharge} = L_{0}+\sum_{\rm i=1}^j L_{\rm i} \cos(\alpha/2)\\
\label{eq:totalwidth}
W_{\rm discharge} = \sum_{\rm i=1}^j L_{\rm i} \sin(\alpha/2),
\end{eqnarray}

where $j$ is the total number of steps the discharge takes 
(where a step is defined as the point at which a new layer of segments
has branched out of the old), so that the total number of segments in
any step is given by $2^j$ and $L_0$ is the length of the initial, solitary segment.

The total number of segments, $N_{\rm segment}$, was also evaluated as
it is related to the total energy. It was calculated by adding the
number of branches over each step,
\begin{equation}
\label{eq:Nseg}
N_{\rm segment} = \sum_{\rm i=0}^{j} 2^i.
\end{equation}
The volume of the discharge can be derived in two ways: a) The
cone volume $V_{\rm cone}$ (Eq.~\ref{eq:vcone}), which is simply the volume
filled by the cone of a height and width
taken from the results of
Eqs.~\ref{eq:totallength} \& \ref{eq:totalwidth} formed by the
discharge branches, and b) $V_{\rm total}$ (Eq.~\ref{eq:vtotal}),
which is the sum over the total number of individual segments of the
discharge, each of which has been treated as a simple cylinder,
\begin{eqnarray} \label{eq:vcone}
V_{\rm tot} = \frac{1}{3} \pi  W_{\rm discharge}^{2}  L_{\rm discharge}\,\,\, [m^3]\\
\label{eq:vtotal}
V_{\rm segments} = \sum_{\rm i=0}^j 2^i  \left (\frac{\pi d_i^2}{4}  L_{\rm i} \right )\,\,\,\,\, [m^3].
\end{eqnarray}
The total dissipated energy, $E_{\rm tot}$, is calculated in a similar
fashion: it is the sum over the total number of segments, where the
length of each segment is multiplied by the energy per length given in
Eq.~\ref{eq:ELrelation},
\begin{equation}
\label{eq:Etot}
E_{\rm tot} = \sum_{i=0}^j  2^i  \left (10^5  L_i \right )\,\,\,\, [J ].
\end{equation}

\section{Results}\label{s:results}

\begin{figure}
\centering
\includegraphics[scale=0.8]{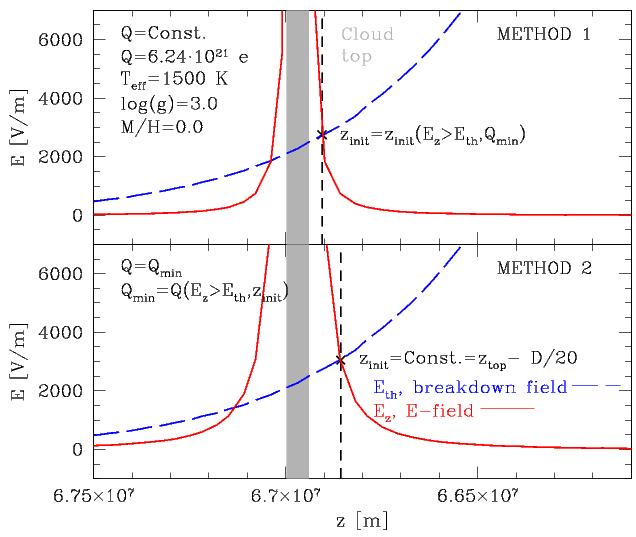}
\caption{Electric fields for the emerging large-scale discharges
  within gas giant and brown dwarf atmospheres (break-down field -- dashed blue line; local electric field -- solid red line). {\bf Top} (case (i)): The total number of charges is prescribed (shown
  for $Q_1=6.24 \times 10^{21}$ e). {\bf Bottom} (case (ii)):
  Atmospheric altitude  of discharge inset is prescribed which allows to determine
  a minimum total charge   needed for the break-down to occur ($Q_{\rm min}=2.58
  \times 10^{22}$ e for the model shown). The {\sc Drift-Phoenix} model atmosphere parameters are $T_{\rm eff}=1500$ K,
  $\log(g)=3.0$, and [M/H]=0.0.}
\label{fig:Emethods}
\end{figure}

In this section, we discuss the results of our model for large-scale
discharges (see Sect.~\ref{ss:lsmodel}) for brown dwarf (BD) and gas
giant (GP) model atmospheres, and for their subsolar ([M/H] = 0.0) and
sub-solar ([M/H] = -3.0) metallicity counterparts.  We investigate how
these scales change with global model parameters like the effective
temperature (T$_{\rm eff}=1500-2000$K).  The range of effective
  temperatures considered comprises those extrasolar, low-mass objects
  where dust clouds form inside the atmosphere and determine the
  observable spectrum.  As outlined in Sect.~\ref{ss:modelansatz},
the results are derived by two different methods:

Case (i) The first method compares the results for each
model atmosphere for a constant total number of charges
($Q_1=6.24 \times 10^{21}$ e and $Q_2 = 3.12 \times 10^{22}$ e)
 Using a prescribed number of charges, the maximum distance from the
 cloud top of a possible discharge initiation within the cloud was
 found by searching for the point in the cloud at which $E_{\rm
   z,Qconst.} > E_{\rm th}$. This point is $z_{\rm init}$.  The top
 panel in Fig.~\ref{fig:Emethods} shows $z_{\rm init}$ for the {\sc
   Drift-Phoenix} model atmosphere of a giant gas planet (T$_{\rm
   eff}=1500$ K, $\log(g)=3.0$) of initial solar metallicity
 ([M/H]=0.0).

Case (ii) The second method compares the results for the minimum
amount of charge, $Q_{\rm min}$, required for a discharge to occur at
a point $z_{\rm init}$ below the cloud top for each individual model
atmosphere. To evaluate comparable minimum charges for each
atmosphere, a set point of discharge initiation below the cloud top
was put at a distance 1/20th the height of the cloud: $z_{\rm init}=
z_{\rm top} - (\Delta H/20)$, where $\Delta H$ is the height of the
cloud. Discharges would initiate at the bottom of this layer, and the
minimum amount of charge, $Q_{\rm min}$, required to overcome the
breakdown field at that point was found such that $E(z_{\rm
  init},Q_{\rm min}) = E_{\rm th}$. From there, the discharge was
allowed to propagate assuming the value of initial minimum charge was
present in the cloud region. The bottom panel in
Fig.~\ref{fig:Emethods} shows $E(z_{\rm init},Q_{\rm min}) = E_{\rm
  th}$ for the {\sc Drift-Phoenix} model atmosphere of a giant gas
planet (T$_{\rm eff}=1500$ K, $\log(g)=3.0$) of initial solar
metallicity ([M/H]=0.0).

The two-fold evaluation process is necessary because we do not know
{\it a priori} how many charges are available in the atmosphere. For
the same reason, \cite{raizer98, pasko00} and \cite{yair09} assume the
presence of a certain number of charges at the height in their
atmospheres under investigation. In contrast, calculations for the
Earth's atmosphere are somewhat guided by {\it in situ} measurements.

\begin{figure}[htbp]
\centering
\includegraphics[scale=0.8]{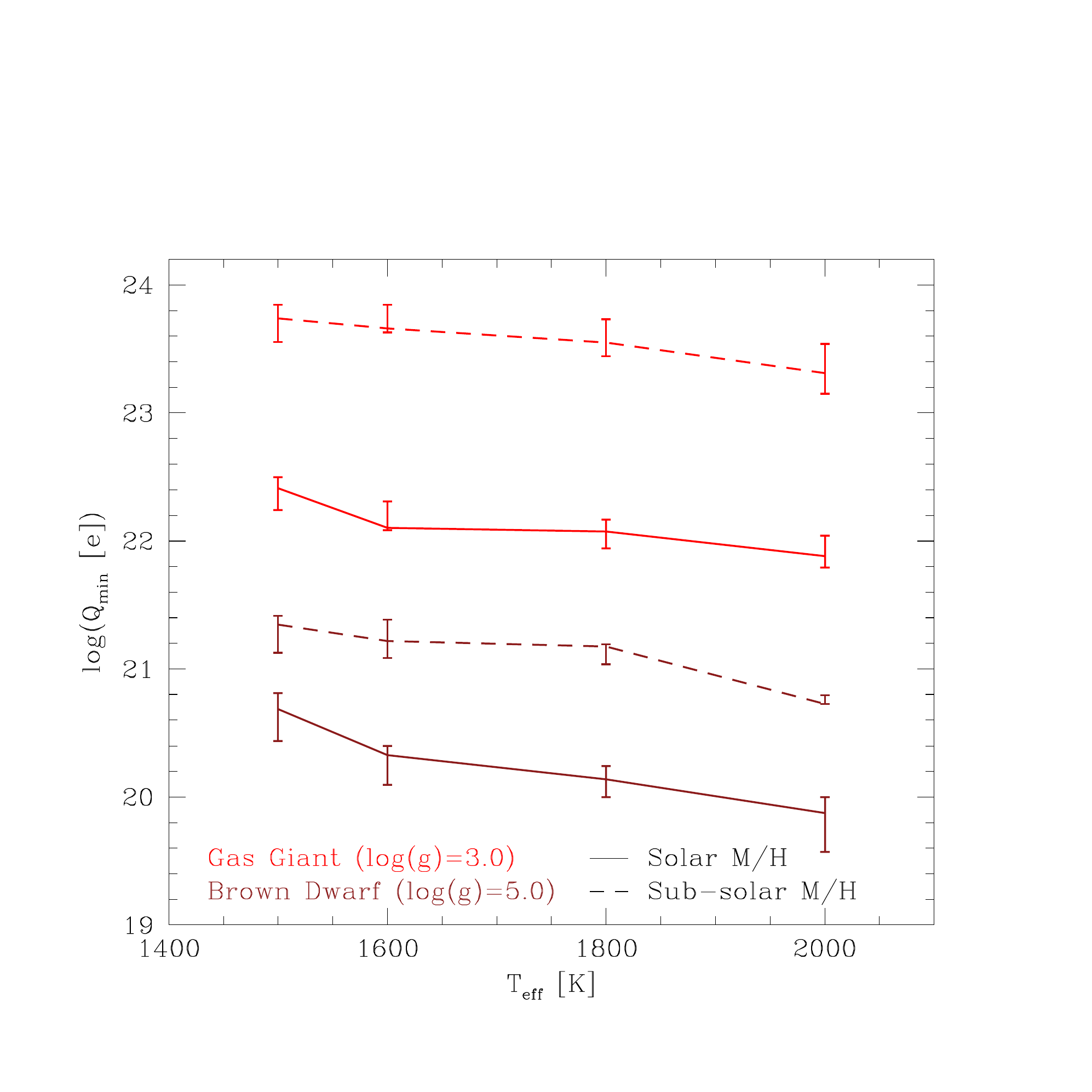}\\*[-1.5cm]
\caption{The minimum charge, $Q_{\rm min}$, required to initiate a
  discharge in model atmospheres of different effective temperatures,
  $T_{\rm eff}$ for BDs (brown) and GGs (red) and for different
  metallicities (solar = solid line, subsolar = dashed
  line). Independent data points are connected to visualise potential
  trends in the results. Due to pressure differences in the cloud
  layers, the brown dwarfs require less amounts of charge than the gas
  giants, and in both cases the lower-metallicity atmosphere models
  require greater amounts of charge than the solar-metallicity
  atmospheres.  The error bars indicate the uncertainty with which
    the cloud height is determined based on the {\sc Drift-Phoenix}
    atmosphere models (see also Sect~\ref{ss:errors}).}
\label{fig:Qmins}
\end{figure}

\subsection{Minimum Charges for electrical breakdown}

We first evaluate the minimum charge required in a cloud layer for a
discharge to initiate just below the cloud top (case (ii), bottom
panel \ref{fig:Emethods}). The results for the different model
atmospheres are shown in Fig. \ref{fig:Qmins}: the amount of charge
required to initiate a discharge decreased from the cool atmospheres
to the hotter atmospheres. This arises as a consequence that the cloud
decks form at lower pressures (hence lower gas temperatures) as the
effective temperature increases.  In analogy with a classical
breakdown in a capacitor discharge, our atmosphere system is operating
in the regime $pd> (pd)_{\rm min}$, where $d$ is the electrode's
separation. This means for a fixed $d$ and a decreasing gas pressure
that the required breakdown voltage, and hence the corresponding
$Q_{\rm min}$, decreases. Therefore, if T$_{\rm eff}$ increases we
expect $Q_{\rm min}$ to decrease.

The breakdown field depends only on gas pressure $p$ and chemical
composition of the gas parametrised by some constants characterising
the discharge.  In this study, we are assuming that H$_2$ is the most
abundant gas species in the atmosphere and so the breakdown conditions
are defined for a H$_2$-dominated gas. However, the chemical
composition of low-metallicity atmospheres will differ to the solar
case. Although H$_2$ remains the most abundant gas-phase species,
other molecules will be less abundant, which has an impact on the
thermodynamic structure of the atmosphere due to radiative transfer
effects: low-metallicity atmospheres are generally more compact for a
given temperature compared to their solar counterparts which causes
the clouds to form at higher pressures (but lower temperature) in
low-metallicity atmospheres. Therefore, the metal abundances have an
indirect influence on $Q_{\rm min}$: we find that $Q_{\rm min}$ is
larger in low-metallicity atmospheres since the clouds form at higher
pressure and require a greater breakdown voltage.


Figure \ref{fig:Qmins} suggests that GP atmospheres require larger
amounts of charge to initiate an electric breakdown of the
gas. However, referring back to Fig.~\ref{fig:pTplot}, we see that the
BD clouds form at higher pressures, which should lead to the BDs
having large values of $Q_{\rm min}$. This discrepancy is caused by
the vertical extension of BD clouds, which have extensions of only
$10^4$ m in comparison to the $10^7$ m clouds in the GP
atmospheres. The small extension causes the field to be larger
throughout the cloud in comparison to the GP case. This leads to BD
models needing lower amounts of charge to initiate a field breakdown
due to the comparatively larger local electric fields.

\begin{figure}[htbp]
\centering
\includegraphics[scale=0.8]{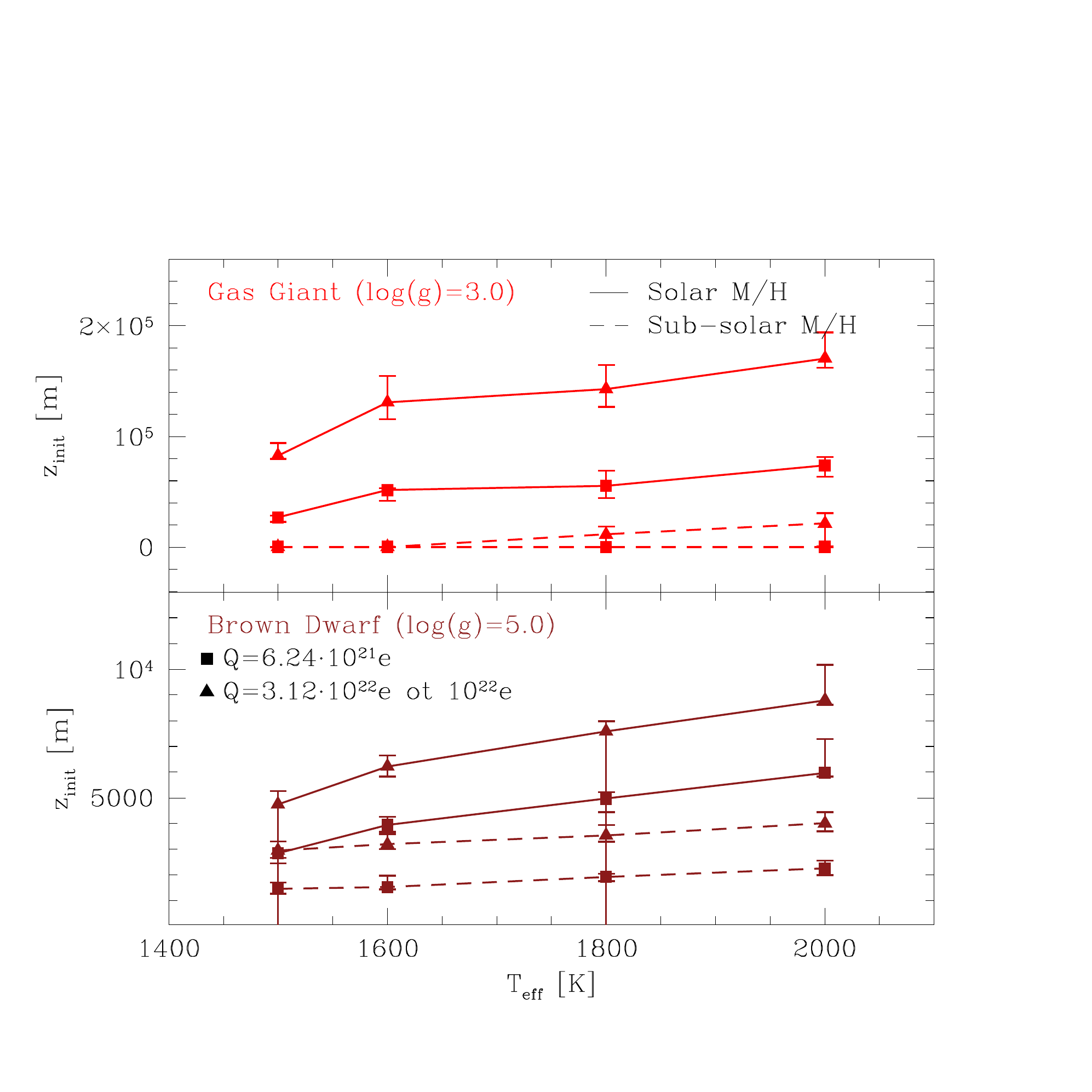}\\*[-1.5cm]
\caption{Distances between the cloud top and the discharge initiation
  height, $z_{\rm init}$, the point at which the local electric field
  grew larger than the breakdown threshold field ($E(z) > E_{\rm th}$)
  for different atmosphere models (top - GPs (log(g)=3.0), bottom -
  BDs (log(g)=5.0),). Results are shown for two different value of
  number of charges (squares: $Q_1=6.24 \times 10^{21}$ e; triangles:
  $Q_2 = 3.12 \times 10^{22}$ e) and for a solar metallicity (solid
  lines) and a sub-solar metallicity (dashed line) case.  The
    error bars indicate the uncertainty with which the cloud height is
    determined based on the {\sc Drift-Phoenix} atmosphere models (see
    also Sect~\ref{ss:errors}.)}
\label{fig:QCZinit}
\end{figure}

\subsection{Large-scale discharge properties}
\label{sec:properties}

This section evaluates the initiation height of the discharge in the
atmosphere; the total length of a discharge event; the total number of
segments that compose the discharge event; the atmospheric volume
affected by the discharge; and the total energy per discharge
dissipated into the atmosphere. All quantities are evaluated for
various sets of {\sc Drift-Phoenix} model atmospheres defined by: the
effective temperature, the gravitational acceleration and the
metallicity ($T_{\rm eff}=1500,1600,1800,2000$ K, $\log(g)=3.0$, 5.0,
$[M/H]=0.0$, -3.0).

\subsubsection{Initiation height}

We evaluate the discharge initiation height, $z_{\rm init}$, for a
given number of charges ($Q_1=6.24 \times 10^{21}$ e and $Q_2 = 3.12
\times 10^{22}$ e) as the height below the cloud top where the local
electric field exceeds the breakdown threshold field, i.e. where $E(z)
> E_{\rm th}$ is satisfied. Results for all the atmospheric types are
plotted in Fig. \ref{fig:QCZinit}, where the two separate point types
represent the two charge amounts. Increasing the prescribed charge
increases the local electric field strength, allowing the local
breakdown field to be overcome in regions of higher pressure. As a
result, since the gas pressure increases with distance below the cloud
top, the distance from the cloud top for a discharge initiation,
z$_{\rm init}$, will increase.  Furthermore, an increase in effective
temperature causes the discharge initiation height to move further
into the cloud, i.e. into regions of higher gas pressure. Cloud
regions in hotter atmospheres form at lower pressures (i.e. higher up
in the atmosphere) with increasing $T_{\rm eff}$ (compare
Fig.~\ref{fig:pTplot}), and the local electric field in these clouds
(for a constant number of charges) is greater than the breakdown field
for a greater distance into the cloud.

The behaviour of the low metallicity GP stands out in these plots as
the initiation distance from the charges is nearly coinciding with the
charge carrying cloud top, where the field is very large. As shown in
Fig. \ref{fig:Qmins}, minimal charges of the order of $10^{23}$ e were
required to initiate a discharge in these low-metallicity atmospheres
due to the cloud tops forming at comparatively higher pressures, hence
deep inside the atmosphere. This means that the only region in which a
discharge could realistically initiate is directly on the charge
carrying surface, where the field is near-infinite. In this scenario,
the dipole electric field model is insufficient to model the system
since the initiation point is in such close proximity to the charged
surface that the spatial distribution of charge and the resulting
field effects would need to be considered. This may suggest that
discharges would not occur in sub-solar metallicity cloud models for
the given number of charges ($Q_1=6.24 \times 10^{21}$ e, $Q_2 = 3.12
\times 10^{22}$ e) assumed in Fig. \ref{fig:QCZinit}.

\subsubsection{Total discharge lengths}\label{ss:tdl}

The total discharge lengths for the two different cases of a
prescribed number of charges (case (i)) and for a calculated minimum
charge, $Q_{\rm min}$, to start a electrical breakdown (case (ii)) are
shown in Figs.~\ref{fig:QCLengths}.
Both cases demonstrate that discharges can be expected to be much more
extended in a brown dwarf atmosphere than in a giant gas planet
atmosphere ($10\times$ the GP values) if both are considered to result
from the same number of charges (left of Fig.~\ref{fig:QCLengths}).

Figure~\ref{fig:QCLengths} shows that a large-scale discharge can
propagate over $0.5~-~4\,$km which is strongly dependent on the
metallicity and surface gravity. Our results suggest that
high-pressure atmospheres, due to high surface gravity or low
metallicity, produce exceptionally large discharges: low-metallicity
atmospheres and brown dwarfs seem to have the largest total discharge
length. The high pressure results in a small mean free path and more
collisions occurring during the electron transit between the cloud
`electrodes'. Each collision randomises the electron motion and will
reduce the electron energy.  Therefore, for an atmosphere with higher
pressure, the magnitude of the electric field required to initiate
electrical breakdown is greater.  Following the empirical scaling
relations, this implies that the initial breakdown potential is
greater and hence the size of the initial diameter of the subsequent
streamer: $d_{\rm init}=n_{V}V_{\rm init}$.  As a consequence, as the
streamer propagates through the atmosphere and begins to branch, the
length of the resulting streamer segments will be greater since
$L\propto d$.  Therefore, we expect discharges to have a greater
spatial scale in high pressure atmospheres such as brown dwarfs.
Furthermore, for a high pressure atmosphere we would
  expect the minimum diameter ($d_{\rm min}\propto T/p$) to be small,
  enabling the streamer to propagate for a greater distance before the
  minimum diameter is reached.

Therefore, a high-pressure gas will allow for the streamer process to
progress over a longer distance through an ambient gas while it will
die out quickly in a low-pressure environment.  Therefore, {\it
  lighting discharges can be expected to be larger, and therefore
  easier to detect, in brown dwarfs and  low-metallicity planetary
  atmospheres.}  Although a larger volume would be affected,
less radiation may be emitted due to a lower number of collisions due
to lower densities, unless a saturated process dominates the emission
process.  The dependence on the effective temperature is not very
strong if $Q = const$ as in Figs.~\ref{fig:QCLengths}.

All of the total discharge propagation lengths
decrease with increasing effective temperature. The total lengths in
the T$_{\rm eff}$=2000 K atmospheres were in some cases 50\% the total
discharge length in a T$_{\rm eff}$=1500 K model atmosphere. The
primary reason is a large geometrical extension of  the cloud in the hotter
atmospheres leads to a lower electric field value for the same amount
charge, which results in a smaller discharge length. A higher number
of total charges led to larger discharge lengths in all model
atmospheres considered.  Comparing the BD model discharge lengths between the $Q = const.$ and
$Q = Q_{min}$ models (Figs.~\ref{fig:QCLengths}) produces a similar conclusion supporting this result:  the minimum number
of charges required to initiate discharges in BD atmospheres were of
the order of $10^{20}-10^{21}$ e, much smaller than the charges
applied in Fig. \ref{fig:QCLengths} (case (i)).


\begin{figure*}
\vspace*{-2cm}
\hspace*{-1cm}
\includegraphics[scale=0.5]{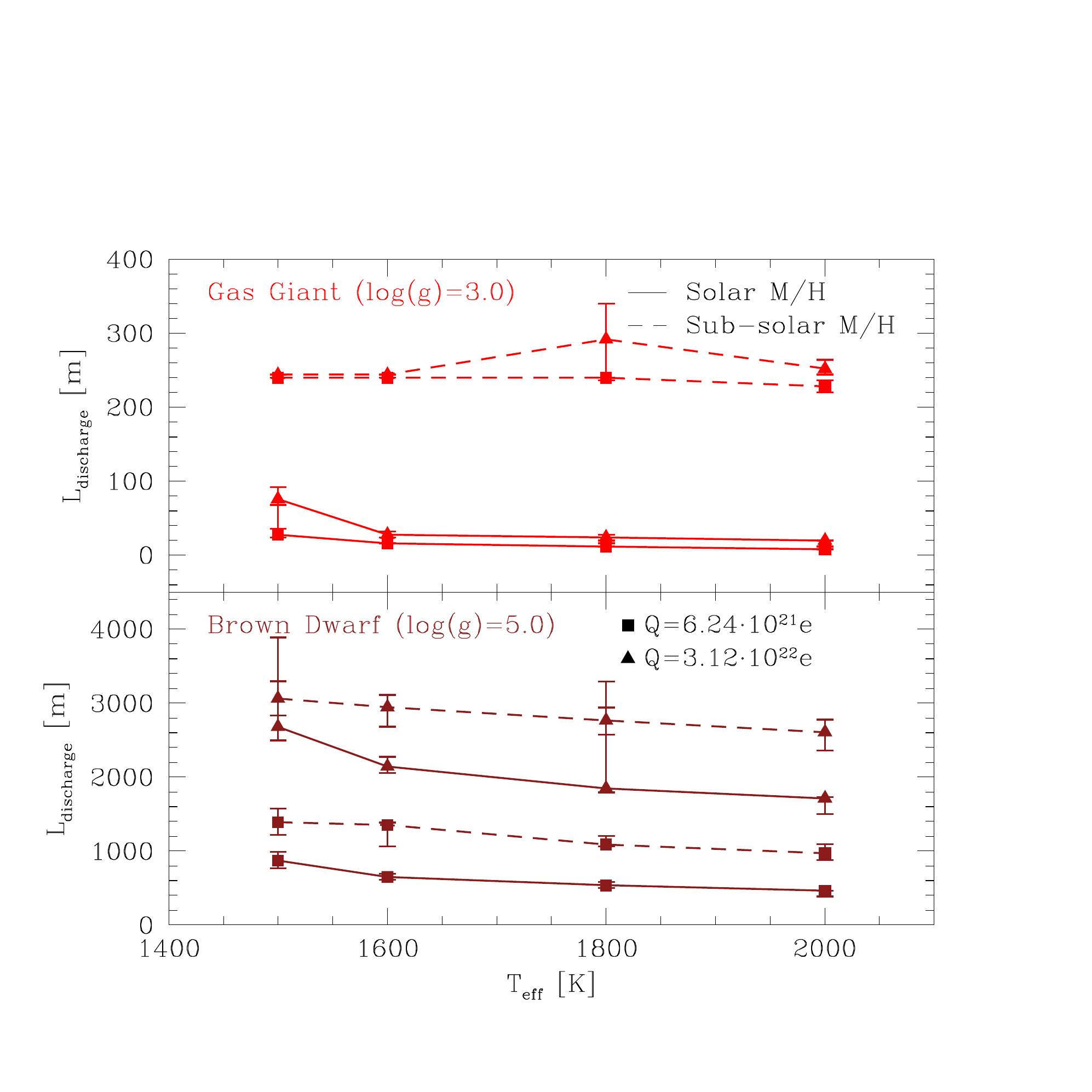}
\hspace*{-2cm}
\includegraphics[scale=0.5]{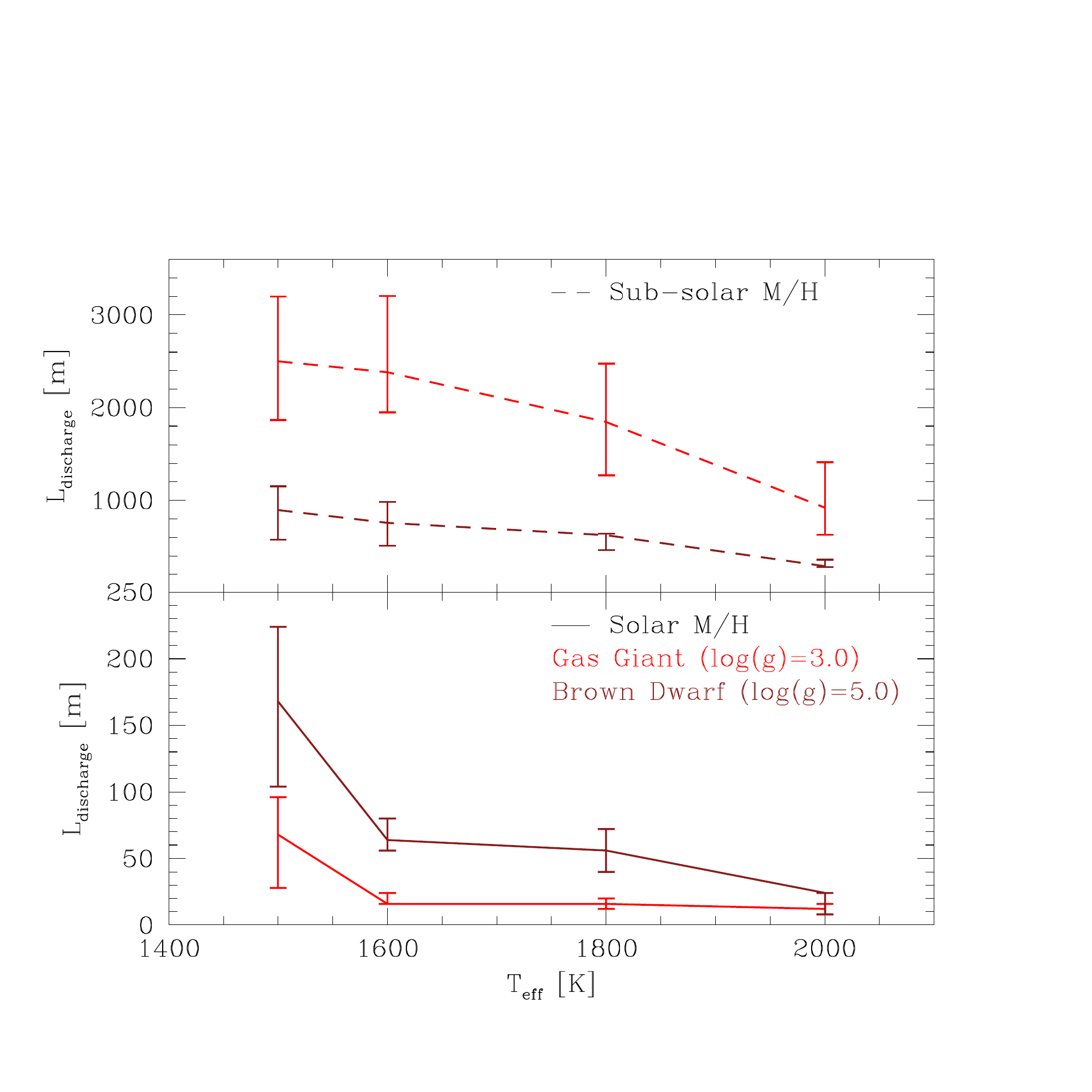}\\*[-2cm]
\caption{The total lengths, $L_{\rm discharge}$, that a large-scale
  discharge can reach in different atmospheres (top - GPs
  (log(g)=3.0), bottom - BDs (log(g)=5.0)) with solar metallicity
  (solid lines) and a sub-solar metallicity (dashed line). {\bf Left:}
  Results for two different value of a constant number of charges
  (squares: $Q_1=6.24 \times 10^{21}$ e; triangles: $Q_2 = 3.12 \times
  10^{22}$ e). {\bf Right:} Results for the minimum number of charges
  needed for a field break-down. The error bars indicate the
    uncertainty with which the cloud height is determined based on the
    {\sc Drift-Phoenix} atmosphere models (see also
    Sect~\ref{ss:errors}).}
\label{fig:QCLengths}
\end{figure*}

\subsubsection{Total number of segments}

We evaluate the the total number of segments (or branches), N$_{\rm
  segments}$ (Eq.~\ref{eq:Nseg}), that compose the whole discharge event in  the
atmosphere.  Figure~\ref{fig:TotBranches} shows N$_{\rm segments}$ for
a prescribed constant number of total charges (top plot) and
 shows N$_{\rm segments}$ for the minimum
number of charges needed for field break-down (bottom plot).  Both figures suggest
an almost exponential decrease of the number of branches across all
atmospheres, and that they are much more numerous in the low-metallicity
atmospheres.

The number of segments grows exponentially with the total length of
the discharge. If a discharge reaches greater lengths, it may branch
more often.  The behaviour of $N_{\rm segments}$ is
  similar to that of the total discharge length: higher pressure
  atmospheres (due to higher gravitational acceleration or lower
  metallicity) require a higher breakdown voltage, resulting in a
  larger $d_{\rm init}$, larger segment lengths $L$ and a lower
  $d_{\rm min}$.  Therefore a large-scale discharge with a greater
  spatial extent is more likely to have a greater number of segments.

\begin{figure*}[htbp]
\vspace*{-2cm}
\hspace*{-1cm}
\includegraphics[scale=0.5]{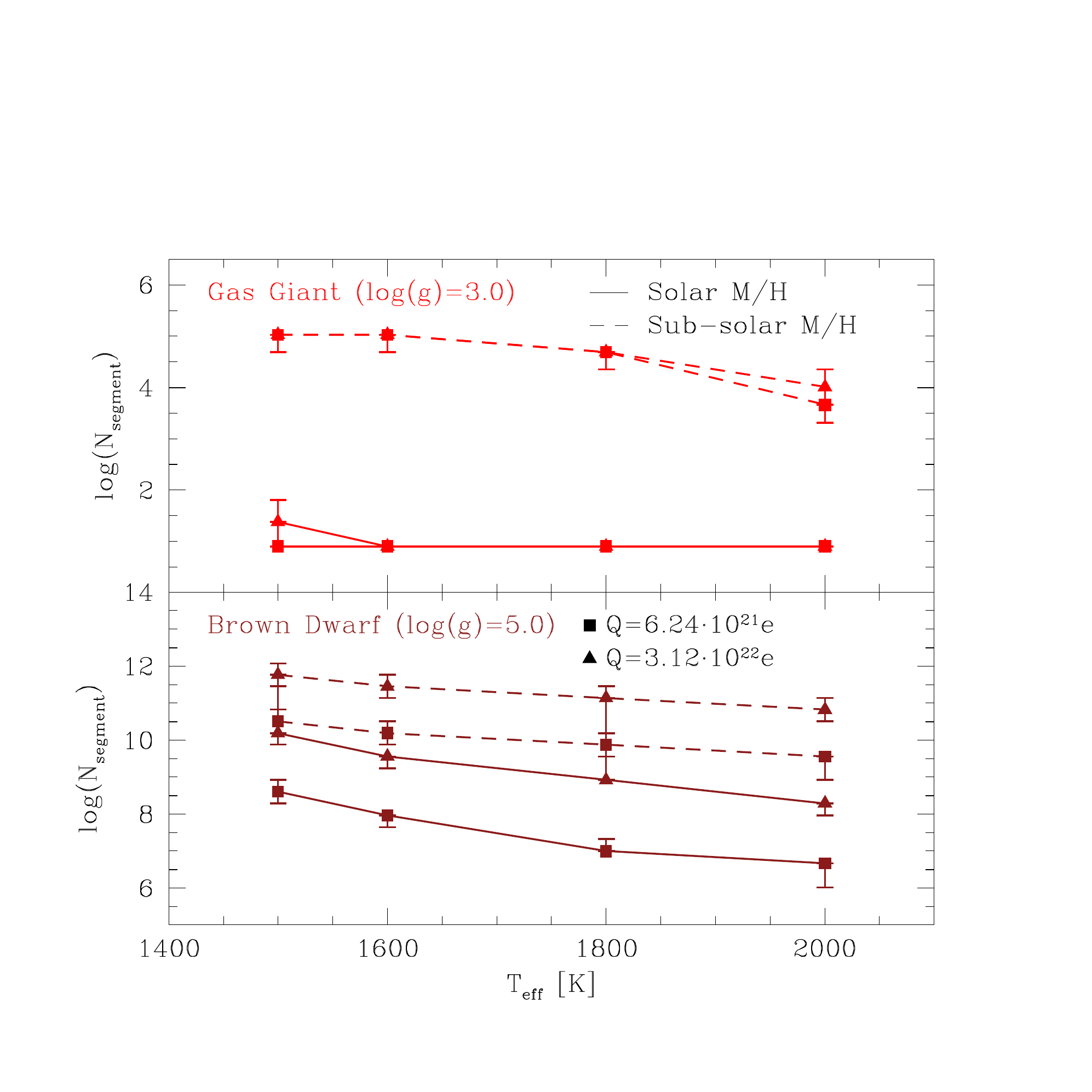}
\hspace*{-2cm}
\includegraphics[scale=0.5]{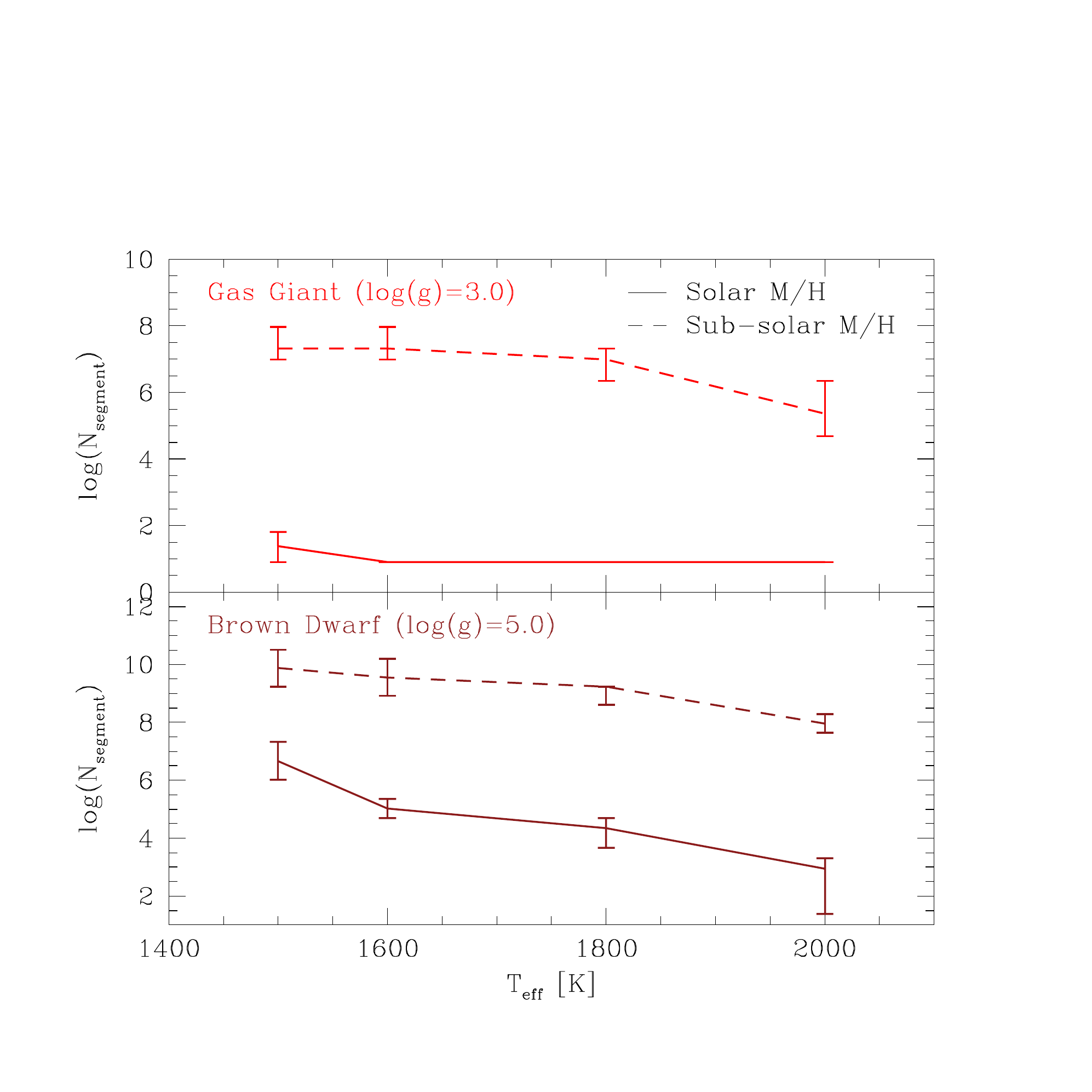}\\*[-2cm]
\caption{The total number of discharge segments, N$_{\rm segments}$,
  in the discharge channel for different model atmospheres
   (top panels - GPs (log(g)=3.0), bottom panels - BDs
  (log(g)=5.0)). All results are shown for  a solar metallicity (solid lines) and
  a sub-solar metallicity (dashed line) case.
{\bf Left:} results for two different value of a constant number
  of charges (squares: $Q_1=6.24 \times 10^{21}$ e; triangles: $Q_2 =
  3.12 \times 10^{22}$ e), {\bf Right:} results for minimum charges needed for field breakdown. The error bars indicate the uncertainty with which the cloud
  height is determined based on the {\sc Drift-Phoenix} atmosphere
  models (see also Sect~\ref{ss:errors}).}
\label{fig:TotBranches}
\end{figure*}

\subsubsection{Total energy dissipated}

We now estimate the total energy dissipated by a large-scale discharge
event in a substellar atmosphere. We utilise the total dissipation
energy per length, E$_{\rm tot}$ in units of
[J/m](Eq.~\ref{eq:ELrelation}), and combine it with our estimate for
the number of segments, N$_{\rm segment}$ (Eq.~\ref{eq:Nseg}), and the
length of each of these segments, L (Eq.~\ref{eq:length}), which leads
to Eq.~\ref{eq:Etot}.  We note, however, that the factor of $10^5$
  in Eq.~\ref{eq:ELrelation} may vary for different atmospheric
  chemistries. Our investigation of the electric break-down conditions in
  \citep{helling13} show, however, that the gas-phase composition does
  only introduce small differences.

\begin{figure*}[htbp]
\vspace*{-2cm}
\hspace*{-1cm}
\includegraphics[scale=0.5]{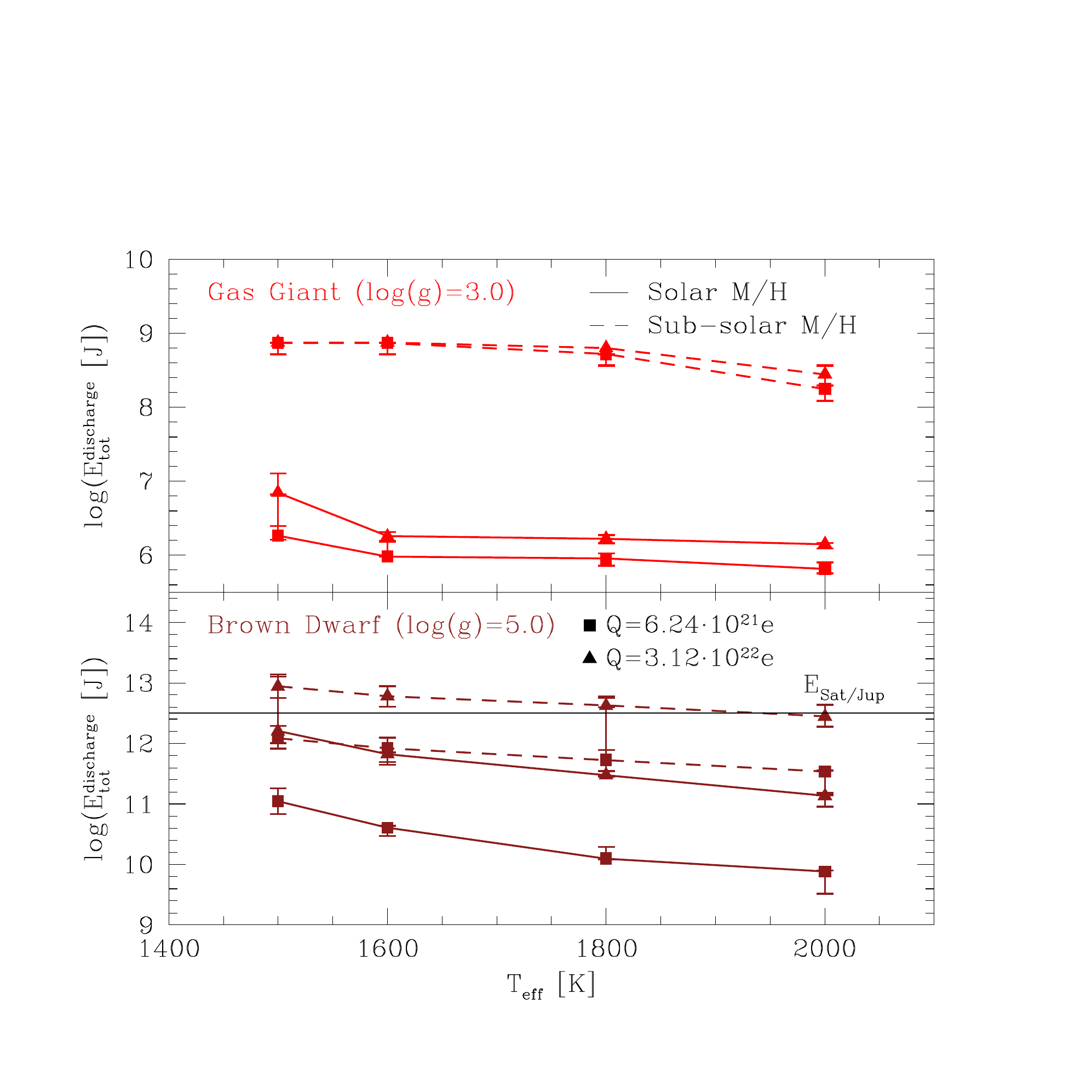}
\hspace*{-2cm}
\includegraphics[scale=0.5]{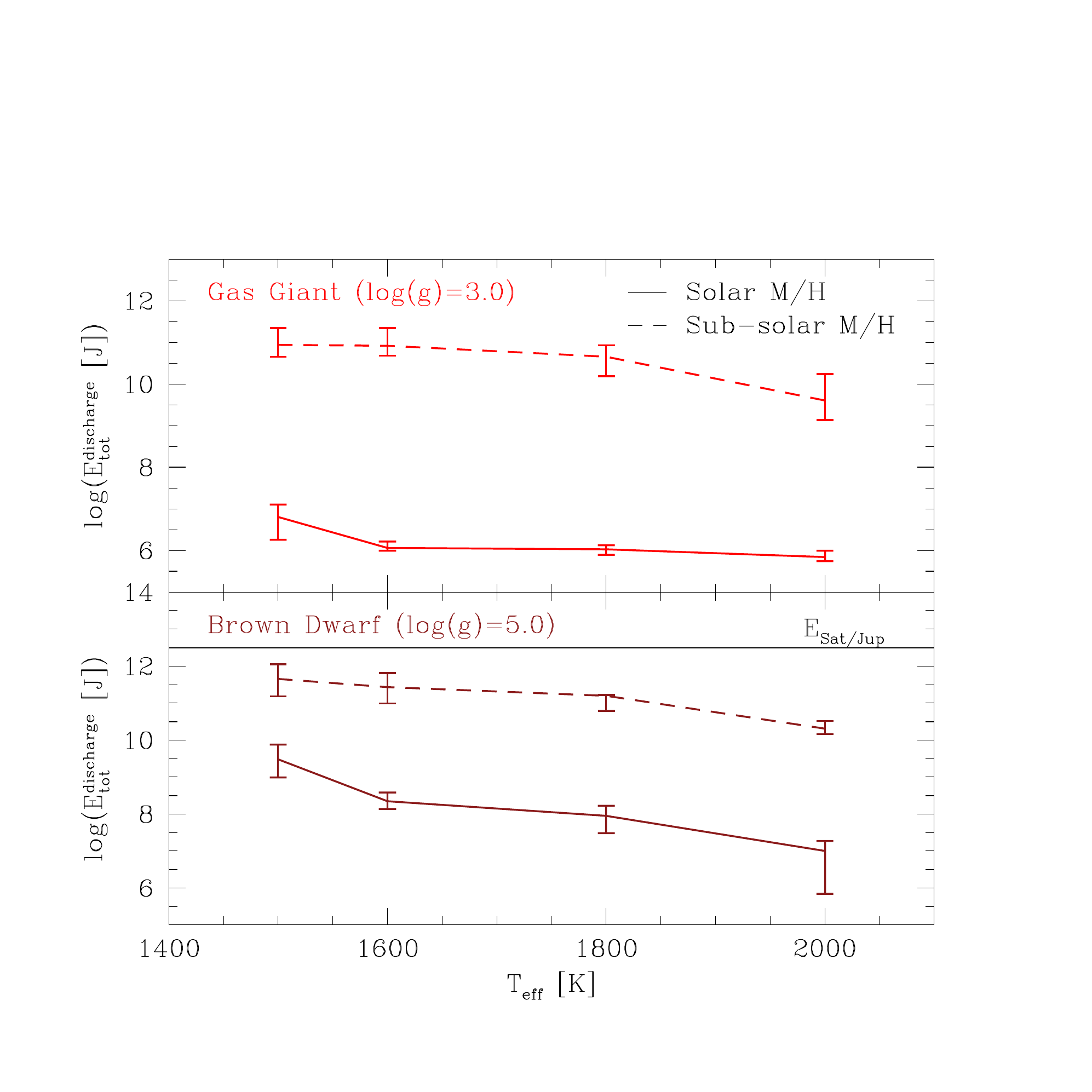}\\*[-2cm]
\caption{The total dissipated energy for different model atmospheres
   (top panels - GPs (log(g)=3.0), bottom panels - BDs
  (log(g)=5.0)). All results are shown for  a solar metallicity (solid lines) and
  a sub-solar metallicity (dashed line) case.
{\bf Left:} results for two different value of a constant number
  of charges (squares: $Q_1=6.24 \times 10^{21}$ e; triangles: $Q_2 =
  3.12 \times 10^{22}$ e), {\bf Right:} results for minimum charges needed for field breakdown. The error bars indicate the uncertainty with which the cloud
  height is determined based on the {\sc Drift-Phoenix} atmosphere
  models (see also Sect~\ref{ss:errors}).} 
\label{fig:QCEnergies}
\end{figure*}

\begin{figure}[htbp]
\centering
\includegraphics[scale=0.7]{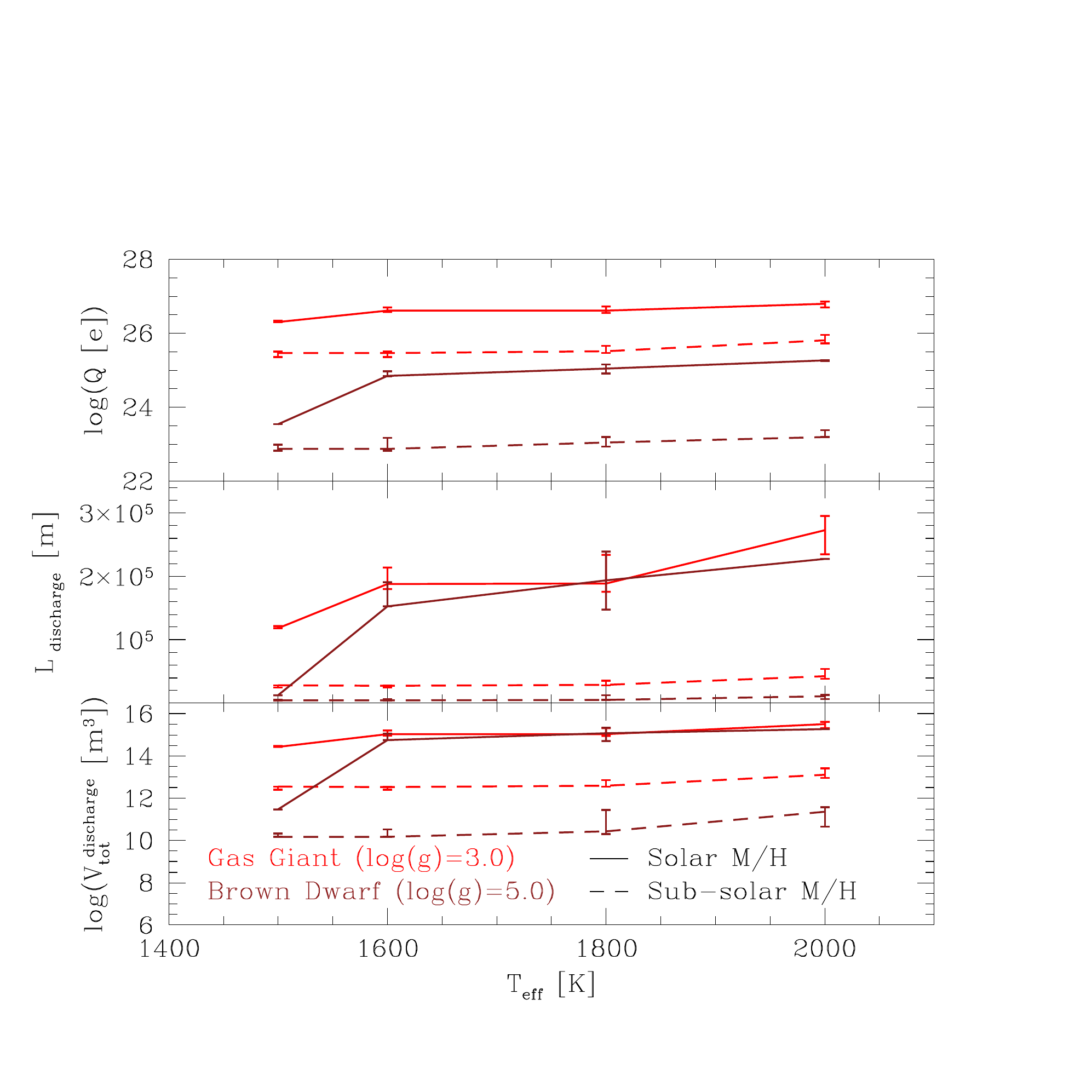}\\*[-1.5cm]
\caption{Characteristic discharge values for a given total dissipation energy, E$_{\rm tot}=10^{13}$J.
{\bf Top:} The minimum charge, Q$_{\rm min}$ [e], needed to
  achieve a field-breakdown with a total dissipation energy E$_{\rm
    tot}=10^{13}$ J (E$_{\rm tot, Jupiter} =
  10^{12}\,\ldots\,10^{13}$J, E$_{\rm tot, Saturn} \approx 10^{12}$J); {\bf Middle:} Total propagation length, L$_{\rm
  discharge}$ [m], of the large-scale discharge dissipating E$_{\rm
  tot}=10^{13}$ J.; {\bf Bottom:} Total atmospheric volume affected by
the propagating discharge that dissipates a Jupiter/Saturn equivalent
of E$_{\rm tot}=10^{13}$ J. All values are shown for different model
atmospheres.  The error bars indicate the uncertainty with which the cloud
  height is determined based on the {\sc Drift-Phoenix} atmosphere
  models (see also Sect~\ref{ss:errors}).}
\label{fig:EThirteen}
\end{figure}

We evaluate the total dissipated energies (Eq.~\ref{eq:Etot})
depending on the model atmosphere parameters T$_{\rm eff}$, log(g),
and on metallicity.  This is done for both cases: (i) for a constant
number of charges; and (ii) the minimum charges for each
atmosphere. The results in Fig.~\ref{fig:QCEnergies} demonstrate that
the total dissipated energy is of the order of 10$^{6}$ - 10$^{9}$ J
for solar metallicity atmospheres. These values are comparable to
typical solar system values which are E$_{\rm tot, Earth} =
10^8~-~10^9$J, E$_{\rm tot, Venus} = 10^9~-~10^{10}$J, E$_{\rm tot,
  Jupiter} = 10^{12}~-~10^{13}$J, and E$_{\rm tot, Saturn} \approx
10^{12}$J. Our estimates suggest that more energy is released in a
brown dwarf atmosphere than in a giant gas planet because of the large
dissipation length. The total dissipation energy in our example GP
atmospheres is generally more comparable to the lightning dissipation
energy on Earth. However, the total dissipated energy reaches its
highest values of $10^{10}-10^{13}$ J for the low-metallicity objects,
which had both higher values of $Q_{\rm min}$ and longer total
discharge lengths than their solar metallicity counterparts, leading
therefore to higher energies.

For a better comparison with known solar system values of energies of
10$^{12}$-10$^{13}$ J in Saturn's and Jupiter's atmospheric
discharges, the discharge propagation model was ran with increasing
applied charge (similar to the constant charge case) until a
Jupiter/Saturn equivalent dissipation energy of E$_{\rm tot}=10^{13}$
J was reached.  The results plotted in Fig.\ref{fig:EThirteen} show
that larger lengths are required in the hotter atmospheres to reach
the same amount of dissipated energy.  This confirms our previous
result for a constant charge, that smaller discharge lengths should
occur in hotter atmospheres compared to cooler atmospheres
(Sect.~\ref{ss:tdl}).

Our results also demonstrate in Fig.~\ref{fig:EThirteen} that a
geometrically larger downward-propagating discharge would be required
in a GP atmosphere than in a brown dwarf in order to achieve the same
dissipation energy in both objects.  Similarly, the discharge length
is smaller in the low-metallicity atmospheres compared to the
solar-metallicity atmospheres.

This later test of finding the discharge properties for a given total
dissipation energy leads us to conclude that our results are
consistent within the framework of our discharge scaling model. Our
model is based on scaling laws derived from laboratory and numerical
experiments on streamers. However, we can not exclude the possibility
that other processes, not quantified by the experimental studies, can
affect the the amount of energy dissipated or the length scale of the
discharge process.

\begin{figure*}[htbp]
\vspace*{-2cm}
\hspace*{-1cm}
\includegraphics[scale=0.5]{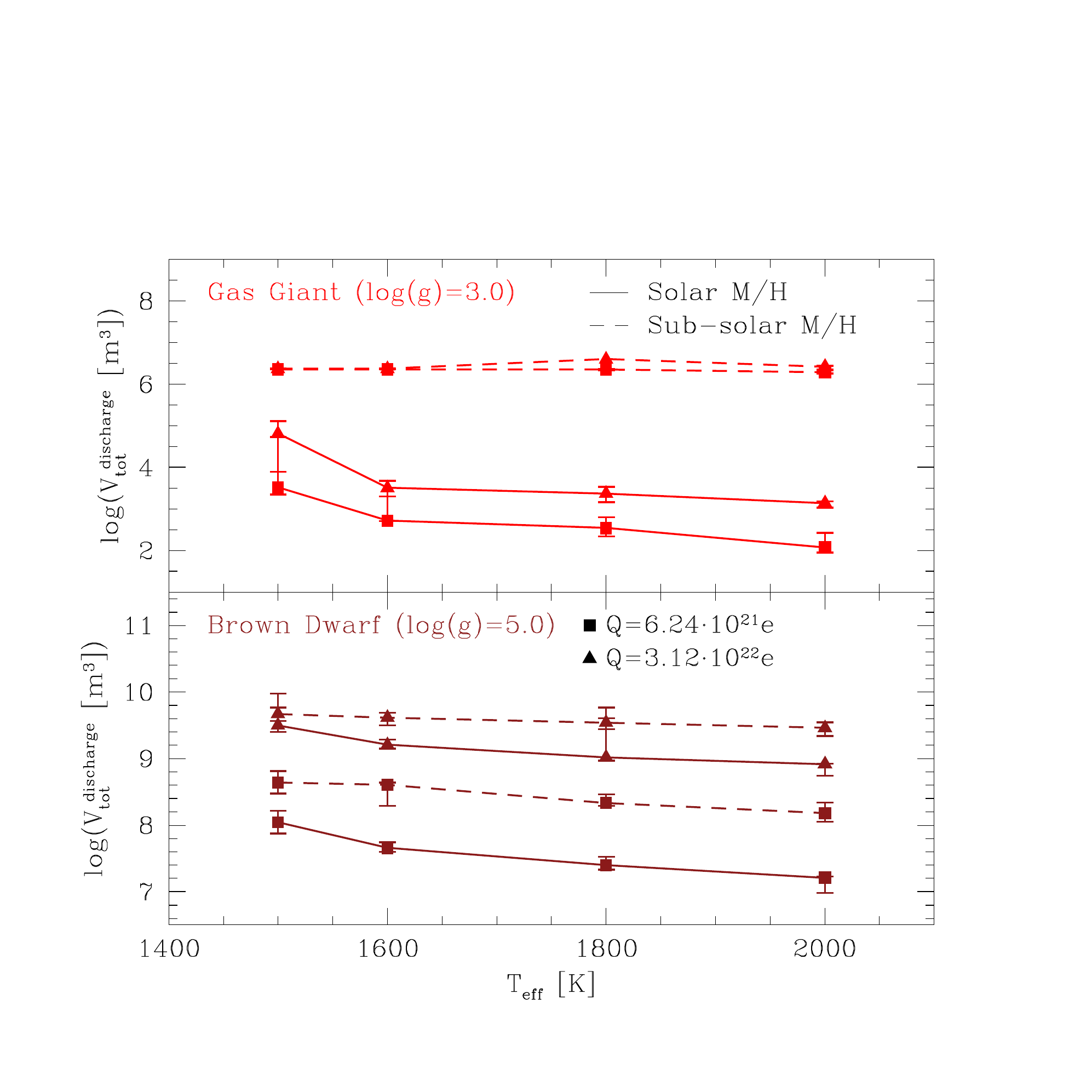}
\hspace*{-2cm}
\includegraphics[scale=0.5]{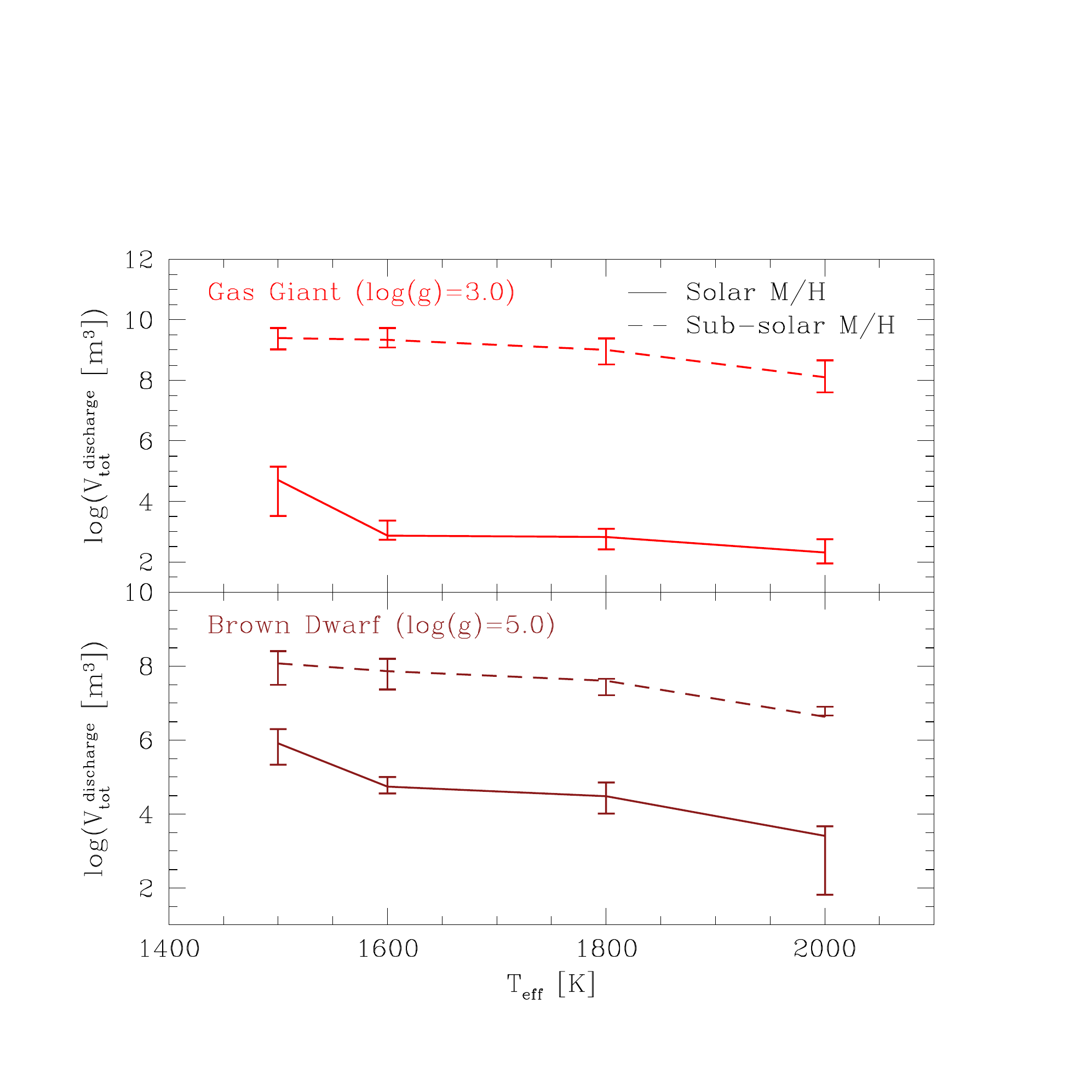}\\*[-2cm]
\caption{The total volume of atmospheric gas that is affected by
  propagating discharges through the atmospheres studied. {\bf Left:} for two
  constant charges ($Q_1=6.24 \times 10^{21}$ e, $Q_2=3.12 \times
  10^{22}$ e), {\bf Right:} the minimum charges. The error bars indicate the uncertainty with which the cloud
  height is determined based on the {\sc Drift-Phoenix} atmosphere
  models (see also Sect~\ref{ss:errors}).}
\label{fig:QCVolumes}
\end{figure*}

\subsubsection{Total discharge volume }

The volumes of atmosphere affected by a discharge propagating through
the atmospheric gas was treated as a cone filled by the discharge
branches as defined in Eqs.~\ref{eq:vcone} and~\ref{eq:vtotal}.  The
results for constant charges are shown in Fig. \ref{fig:QCVolumes}
(left), and those at $Q_{\rm min}$ in Fig.~\ref{fig:QCVolumes}
(right). For discharges of the size of $10^2$ m (GP) and $10^3$ m (BD,
see Fig. \ref{fig:QCLengths}), we observe total cone volumes of the
orders of $10^{4}~-~10^{6}$ m$^3$ and $10^{8}~-~10^{10}$ m$^3$
respectively.  This is the estimated volume of the atmospheric gas
where a population of ions, metastables and electrons has been
injected in the streamer wake. Combined with the local electric field
and associated change in local temperature, this will allow chemical
reactions not normally permitted in the ambient atmosphere.

However, the volume of the atmospheric gas that is exposed to the
discharge may be underestimated in our simplistic scaling model. For
example, a fractal ansatz for discharge propagation as suggested in
\citep{pasko00} might yield larger volume values.

\subsection{Sprites}

Streamer discharges are suggested to determine the early stages of
lightning discharges and of sprites (\citet{phelps74,raizer91,gorm98,
  briels08a}). Sprites, which are massive discharges that occur above
thunderstorms milliseconds after powerful lightning strikes (and are
therefore also referred to as {\it above-cloud discharges}), have a
similar filamentary structure to streamers. It has therefore been
suggested (e.g. \cite{briels06}~and~\cite{ebert10}) that streamers and
sprites share similar underlying physical mechanisms. Massive {\it
  National Lightning Detection Networks} provide evidence that sprites
and lightning discharges are linked: About 80\% of the observed
sprites on Earth coincide with lightning ground strokes
\citep{bocci95}. This is confirmed by numerical modelling in
combination with high-speed measurements of sprite optical emissions
in \cite{liu09} and \cite{gam11}, as well as by dedicated observation
campaigns for single events (\cite{fuellekrug20013})

An electromagnetic pulse that results from a very large cloud
discharge in the Earth's atmosphere can transfer (positive) charges
downwards. The consequence is a large electrostatic field above the
thundercloud that exceeds the (classical) threshold electric field for
breakdown and creates an upward directed sprite discharge
(e.g. \citet{ryha2012}). The classical breakdown field
(Sect.\ref{ss:Eth}) does generally not incorporate the idea of a
runaway breakdown as described in e.g.  \cite{roussel08}, and
therefore overestimates the critical field strength needed for
electrical breakdown to start.  However, the classical approach can
still provide guidance for first-order-investigations as performed in
this paper in order to gain insight into how sprite extensions may
change in different, extrasolar environments.

The comparison of the local electric field resulting from a
large-scale charge distribution with the critical (classical)
break-down field (Fig.~\ref{fig:SpriteE}) shows that the local
electric field can exceed the breakdown field below and above the
charge-carrying cloud layer. This indicates that the discharge process
can start downwards into the cloud and along a positive density
gradient {\it and} upwards above the cloud travelling into a negative
density gradient.  The downwards travelling discharges could be
considered equivalent to intra-cloud lightning on Earth, and the
upward directed discharge resulting in an upward travelling ionisation
front could be considered equivalent to a sprite.

\begin{figure}[htbp]
\vspace*{-2cm}
\hspace*{-1cm}
\includegraphics[scale=0.5]{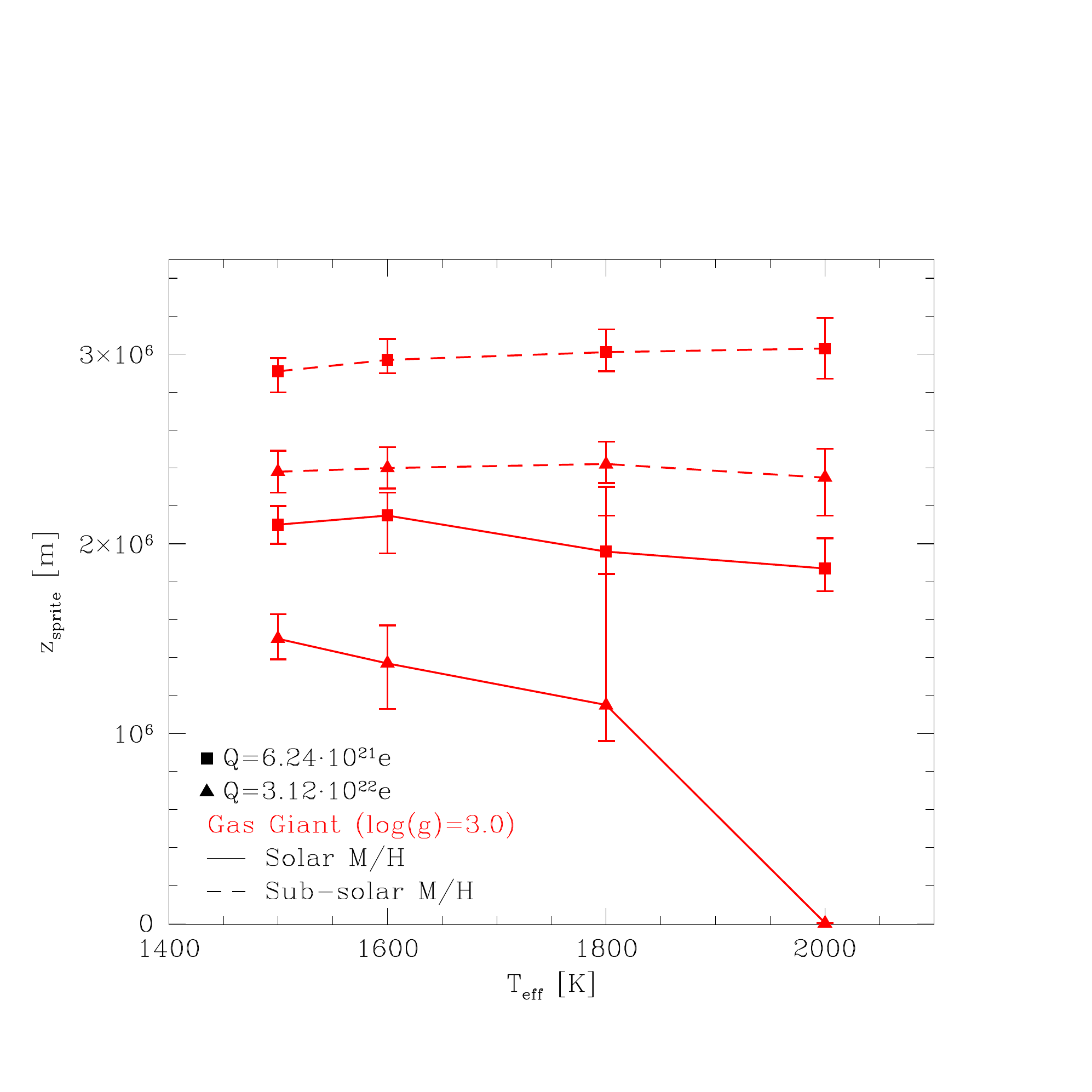}
\hspace*{-1cm}
\includegraphics[scale=0.5]{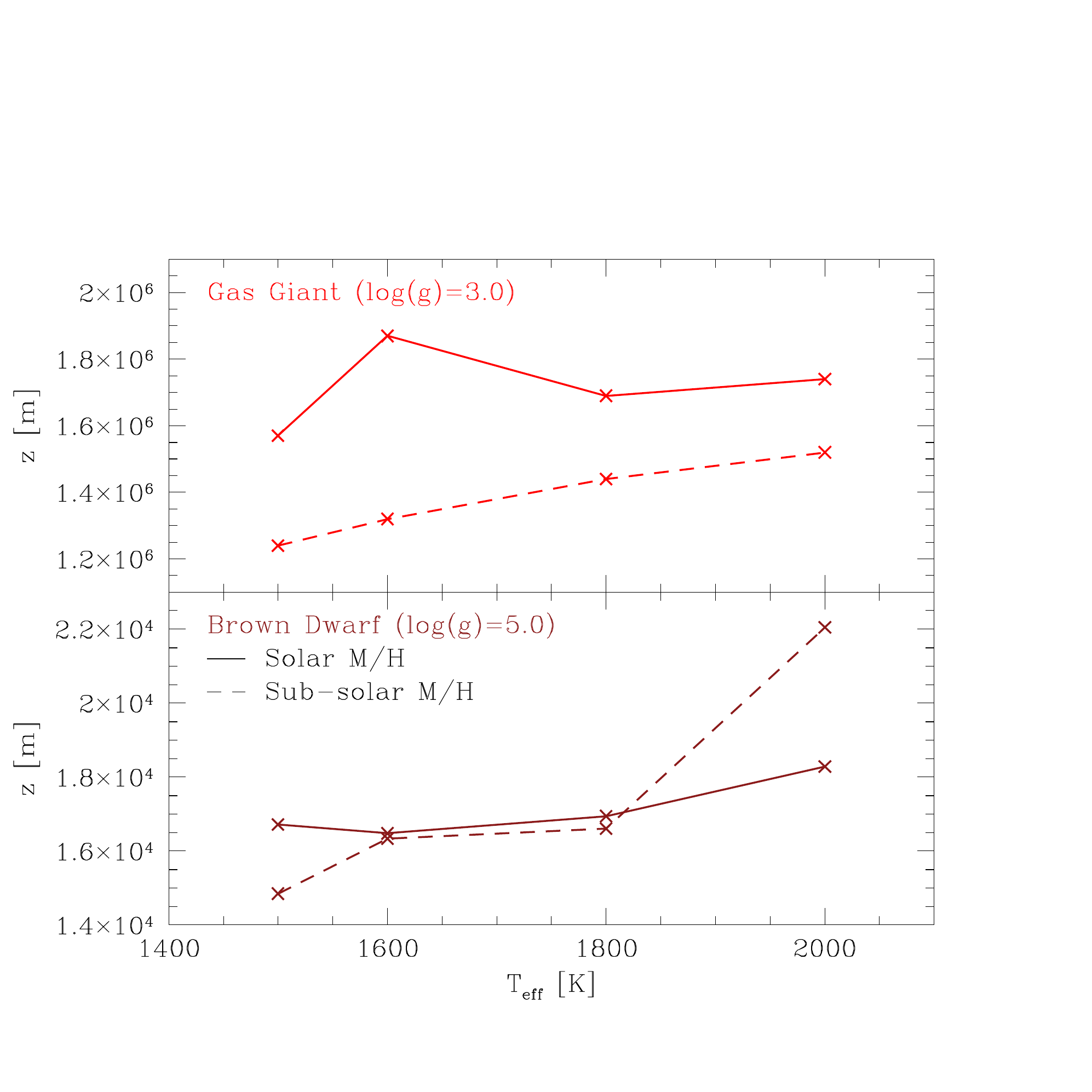}\\*[-2cm]
\caption{The heights above the cloud top for possible sprite
  initiation points for different model atmospheres (red - GPs
  (log(g)=3.0), brown - BDs (log(g)=5.0)). All results are
  shown for a solar metallicity (solid lines) and a sub-solar
  metallicity (dashed line) case.  {\bf Left:} results for two
  different value of a constant number of charges (squares: $Q_1=6.24
  \times 10^{21}$ e; triangles: $Q_2 = 3.12 \times 10^{22}$ e), {\bf
    Right:} results for minimum charges needed for field breakdown. The error bars indicate the uncertainty with which the cloud
  height is determined based on the {\sc Drift-Phoenix} atmosphere
  models (see also Sect~\ref{ss:errors}).}
\label{fig:QCSprite}
\end{figure}

The electric field for a possible sprite, as shown in
Fig. \ref{fig:SpriteE}, was evaluated in the regions above the cloud
deck in the different cloud-forming model atmospheres considered
here. Figure~\ref{fig:QCSprite} shows at which height above a cloud
top a sprite would initiate for brown dwarfs and giant gas planets of
different $T_{\rm eff}$. The condition for the appearance of sprites is
the same as that for other discharges considered here, namely that the
electric field above the cloud charge distribution must be larger than the breakdown
field. Figure~\ref{fig:QCSprite} therefore plots at which height above
the cloud $E_{\rm init, sprite} > E_{\rm th}$. This method was used by
\citet{yair09} to investigate the possibility of Sprites in other solar system planets than Earth.

Figure~\ref{fig:QCSprite} (left) only contains results for the models
describing atmospheres of giant gas planets because the electric field
strength exceeds $E_{\rm th}$ for all locations above the cloud in
brown dwarf atmospheres. Hence, sprites would potentially appear at
any height above the cloud layer in a brown dwarf.  A similar
behaviour occurs in giant gas planet atmospheres
(Fig.~\ref{fig:QCSprite}, left) where for an increasing number of
charges, hence an increasing electric field strength, sprites appear
closer and closer to the cloud top. The distance from the cloud top
dropping to zero in the solar metallicity $T_{\rm eff}=2000$ K
atmosphere signifies the same result as that for the BDs: that a
sprite could occur at any point above the cloud.

A less straight forward behaviour occurs if the minimum charges for
field break-down is considered (Fig.~\ref{fig:QCSprite}, right): the
distance from the cloud top increases slightly as the effective
temperature rises.

The total gas volume affected by one streamer should intuitively be
larger than a traditional lightning-affected volume because a Sprite
can travel much further.  However, it is not obvious how the column
density would differ between a lightning and a Sprite discharge
because of the outward negative density gradient in an atmosphere. In
the framework of our streamer propagation model, we can not evaluate
the atmospheric gas volume affected by a sprite as the diameter of the
discharge channels increase as they propagate into lower pressure
regions. Our method diverges for sprites (Eq.~\ref{eq:dmin}) because
we utilise the existence of a minimum streamer diameter as termination
criterion for streamer propagation. This is appropriate for streamer
propagation along a positive pressure gradient into an atmosphere. If
the streamer propagates along a negative pressure gradient, its
diameter increases and its propagation would be terminated by the
increasing mean free path of the avalanche electrons which at some
point will not have enough energy to travel further. However, this
process is not incorporated by our simple streamer propagation model.

\section{Uncertainty assessment of large-scale discharge properties}

\subsection{The influence of the gas composition}
\label{sec:diffatmo}

All of the above results were calculated for {\sc Drift-Phoenix} model
atmospheres using the parametrisation for the breakdown field $E_{\rm
  th}$, for the chemical composition of a Jupiter atmosphere
\citep{helling13}. This is inconsistent in comparison to the initial
solar or subsolar element abundances used in the atmosphere models. We
therefore assess in how far different chemical compositions of the
atmospheric ionising gas may change our results.


First, we calculate the minimum charges, $Q_{\rm min}$, needed to
achieve a local electric field larger than the threshold breakdown
field for the different atmospheres, {\it assuming} an atmospheric gas that
has a composition comparable to Earth, Mars, Venus, Saturn (parameters
used from \citep{helling13}). The results are plotted in
Fig. \ref{fig:DiffQmins} for giant gas planet atmospheres
($\log(g)=3.0$) of solar composition ($[M/H]=0.0$) and for different
effective temperatures. The charge needed to initiate electrical
breakdown is largest in an Earth-like N$_2$-dominated atmospheric gas;
and smallest in a atmosphere of a Jupiter-like H$_2$-dominated
composition. The results may reflect the higher ionisation energy of
N$_2$ (15.5808 eV) compared to H$_2$ (15.4259eV). However, the
difference between these two values is not very large, which leads us
to refer to the effect of so-called Penning-mixtures (or neon lamp
effect). The effect here is that the gas contains a species which is
easier to ionise than the most abundant species; hence, it efficiently
seeds the field breakdown at lower voltages.  In general, the numbers
are not significantly different between the different atmospheric
gases.

\begin{figure*}[htbp]
\vspace*{-2cm}
\hspace*{-0.5cm}
\includegraphics[scale=0.5]{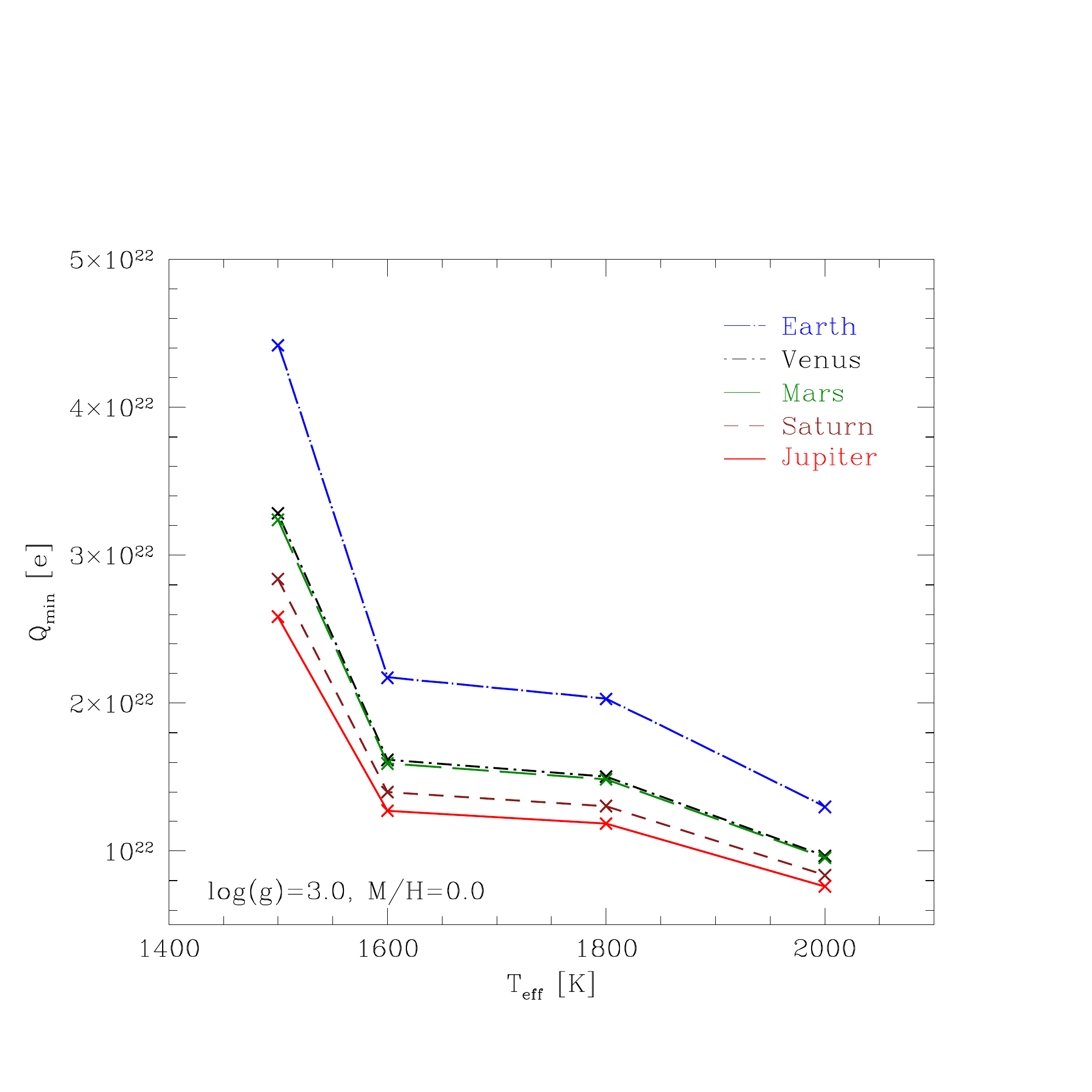}
\hspace*{-2cm}
\includegraphics[scale=0.5]{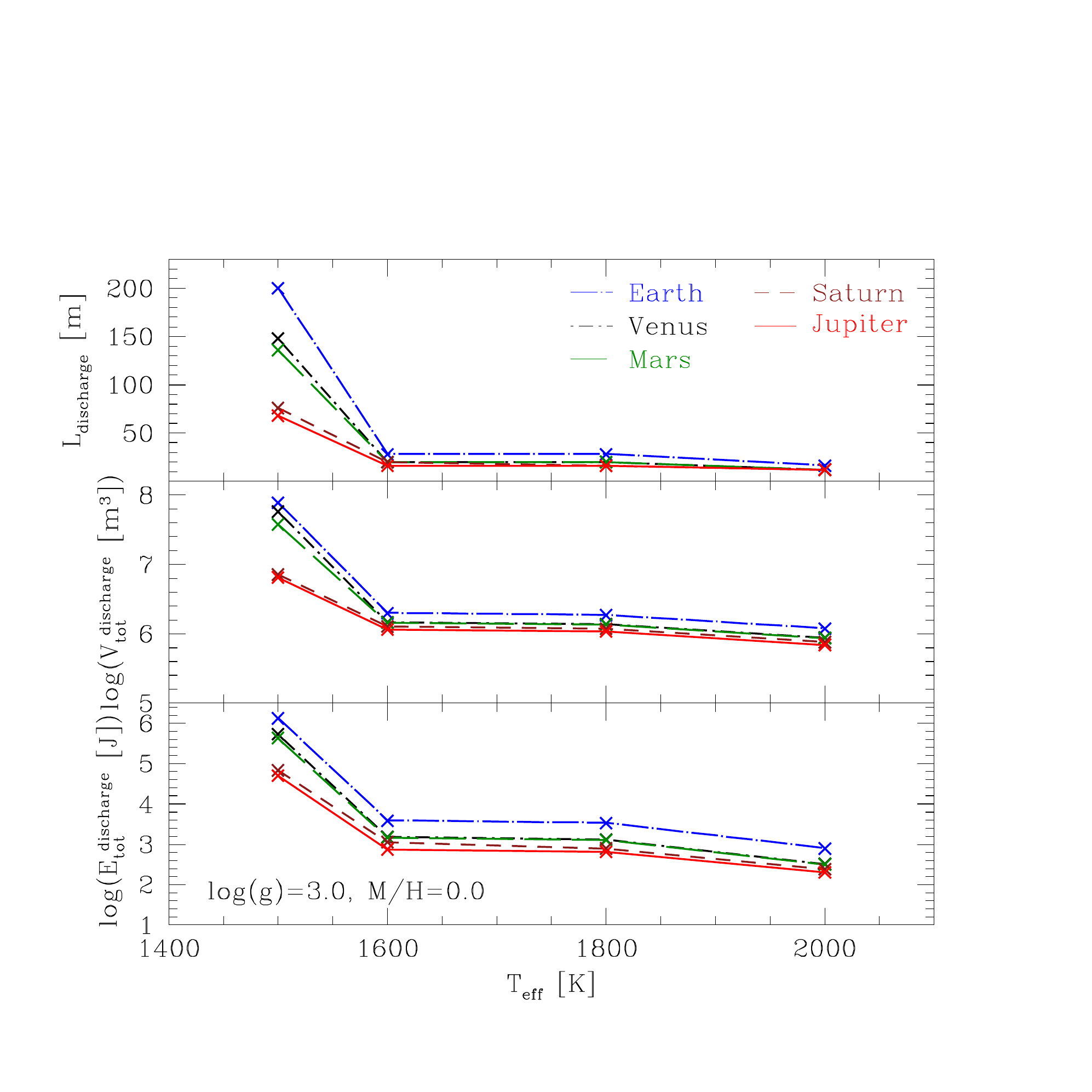}\\*[-1.5cm]
\caption{Dependence of large-scale discharge properties on chemical
  composition  that influences the break-down values of atmospheric gas
  for different T$_{\rm eff}$. {\bf Left:} The minimum charge, $Q_{\rm
    min}$, required to overcome the breakdown field at every point in
  the atmosphere for ionising gases of different chemical composition.
  {\bf Right:} The total discharge length $L_{\rm discharge}$ (top),
  the total volume $V_{\rm tot}^{\rm discharge}$ (centre), and the
  total energy $E_{\rm tot}^{\rm discharge}$ (bottom) are plotted for
  different discharge chemistries with each minimum charge as shown on
  the left of this figure.  Venus and Mars lie very close. We evaluate
  giant gas planet atmospheres ($\log(g)=3.0$) of different T$_{\rm
    eff}$ and solar metallicity ($[M/H]=0.0$).}
\label{fig:DiffQmins}
\end{figure*}

Using these minimum charges in Fig. \ref{fig:DiffQmins}, we show what
effect different atmospheric chemistries might have on the total
propagation length of the discharge (total discharge length),
$L_{\rm discharge}$; the total atmospheric volume affected by the
discharge ionisation, $V_{\rm tot}^{\rm  discharge}$; and the total energy
dissipated into the surrounding gas, $E_{\rm tot}^{\rm discharge}$
(left of Fig.~\ref{fig:DiffQmins}).

The total propagation length of the whole discharge event does not
change appreciably with the chemical composition of the gas for the
higher-temperature models. Given the differences in $L_{\rm
  discharge}$ and the total atmospheric volume affected by the
discharge ionisation, $V_{\rm tot}^{\rm discharge}$, it is only
logical that the total dissipation energy is higher in an Earth-like
atmosphere compared to Jupiter.  The effect of the different
atmospheric chemical compositions appear to have the largest impact on
the discharge properties of the low-temperature models.

\begin{figure}[htbp]
\centering
\includegraphics[scale=0.6]{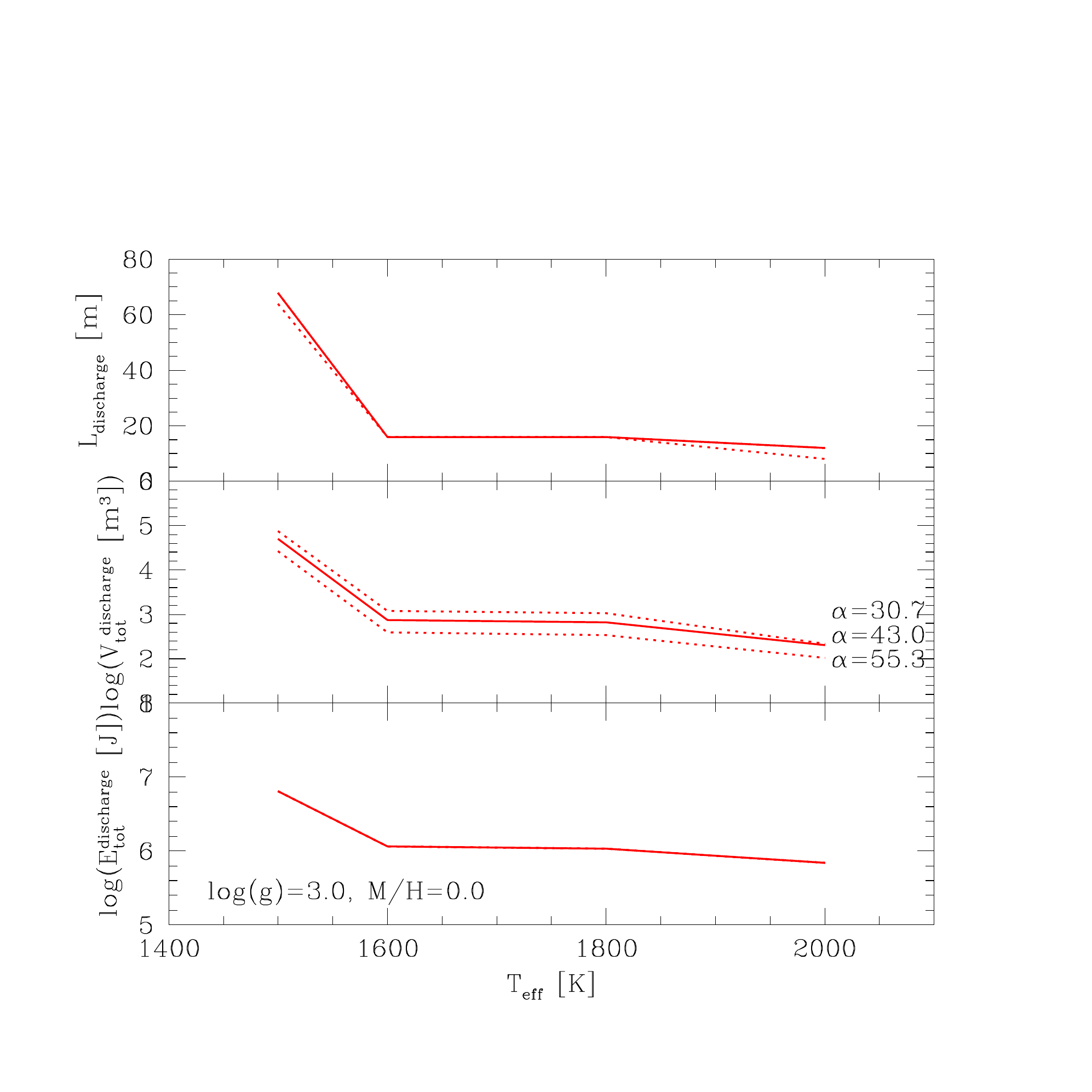}\\*[-1.5cm]
\caption{Dependence of large-scale discharge properties on different
  branching angles. {\bf Top: } discharge lengths, {\bf Middle:}
  volume of the discharge cone, {\bf Bottom:} total dissipated
  energies. The dotted lines represent the upper and lower limits of
  the branching angle \citep{nijdam08}. The results are for
  atmospheres with $\log (g)=3.0$, $[M/H]=0.0$.}
\label{fig:Angles}
\end{figure}

\subsection{The influence of experimental uncertainties}
\label{ss:errors}

All our estimates in the previous section are based on laboratory
experiments and evaluation of numerical results. We assess the
uncertainties introduced by the somewhat large uncertainties in the
branching angle, $\alpha$, and by the determination of the cloud
boundaries from the {\sc Drift-Phoenix} model atmospheres.  

{\it Branching angle $\alpha$:} The branching angle was given by
\citet{nijdam08} with an error estimate of $\alpha=43.0 \pm
12.3^{\circ}$ (Sect.~\ref{sec:scalinglaws}). Figure \ref{fig:Angles}
demonstrates how this uncertainty affects the values for total
lengths, energies and volumes using the minimum
charge model, $Q_{\rm min}$. Each upper and lower limit for each  $\alpha$ is plotted. The
total dissipation energy is not affected; the total discharge length
show small  variations; and  the total volume affected by the
discharge shows an uncertainty of approximately  half an order of magnitude.

{\it Cloud boundary:} Cloud boundaries in the {\sc Drift-Phoenix}
atmosphere models were derived by looking at the nucleation maximum
that defines the cloud top, and a lower and upper cloud boundary was
defined for each. In the gas giants, the  corresponding error spans
$\approx10^5$ m; and in the brown dwarfs
these errors were of the order of $\approx10^3$ m due to the smaller
cloud sizes. The program was run for the same conditions as the
original points, for each upper and lower limit to find the
difference.  These limits are represented by the error bars on the
plots.  (Figs.\,~\ref{fig:Qmins}-~\ref{fig:QCSprite} (left)). Although
these errors can be large, the principle findings of our study do not
change. Moreover, the resulting errors on the total dissipation energy
(Fig.~\ref{fig:QCEnergies}) and the total volume affected by one
large-scale discharge event (Fig.~\ref{fig:EThirteen}) are small.

\section{The effect of electrical discharge events on the local gas chemistry}\label{s:chem}

Atmospheric electric discharge events, and associated physicochemical
interactions, are highly non-equilibrium processes. The accelerated
electrons, those of the streamer, ionise the ambient medium creating a
significant population of ions, radicals, metastable species and
additional electrons.~~Driven by the prevalent local electric field
permeating the medium, chemical reactions are allowed that would
normally be forbidden in a solely thermally-driven system. Such
non-equilibrium plasma chemistry is complex and formidable to model.
To simplify matters, we focused our attention to timescales where the
generated plasma species was extinguished, which left us with a heated
volume of thermalised gas as result of the discharge event. As the gas
cooled, we looked at the subsequent quasi-static equilibrium states and
the resulting chemistry as a function of time.

We present a tentative, initial discussion of the effect that energy
dissipated by a discharge event (in a discharge channel) has on the
atmosphere, the atmospheric chemistry and the local gas-phase
composition. We assume that the gas-phase chemistry remains in steady
state, hence, all gas-phase kinetics proceed on time-scales shorter
than $\tau_{\rm chem}=10^{-4}$s. However, this may be incorrect for
some of the species considered as outlined in e.g. \cite{lorenz08}.
Similar attempts have been made to investigate the atmospheric
chemistry of the Earth and other solar-system planets. For example,
\citet{kovacs10} investigated the evolution of Titan's chemistry as a
result of lightning strikes.  \citet{kovacs10} considered a lightning
channel gas cylinder of diameter 0.025 m and of initial maximum
temperature 30000 K.  The plasma in the lightning column extinguished
quite rapidly, but the gas temperature decreased on a longer time
scale. This intense heating of the lighting channel is observed as
optical emission in Earth and on Jupiter
(e.g. \citealt{zarka04}). \citet{kovacs10} numerically solved the heat
conductivity equation in order to calculate the time-dependent
temperature cooling profile of the lightning channel (their
Fig. 1). Titan has a different chemical composition than the
extrasolar objects considered in this paper; however, the principle
physical mechanisms of an electrical breakdown are the same,
independent of the local difference in chemistry.  Foreseeable
  differences in time-scales can result from the chemistry dependent
  cooling efficiency, particularly in rich gas mixtures (e.g Fig. 12
  in \citealt{woitke96}). Such time-scale effects  are not too important for our study
  as our main interest is the changing temperature during the cooling
  process.

\begin{figure}[htbp]
\centering
\includegraphics[scale=0.8]{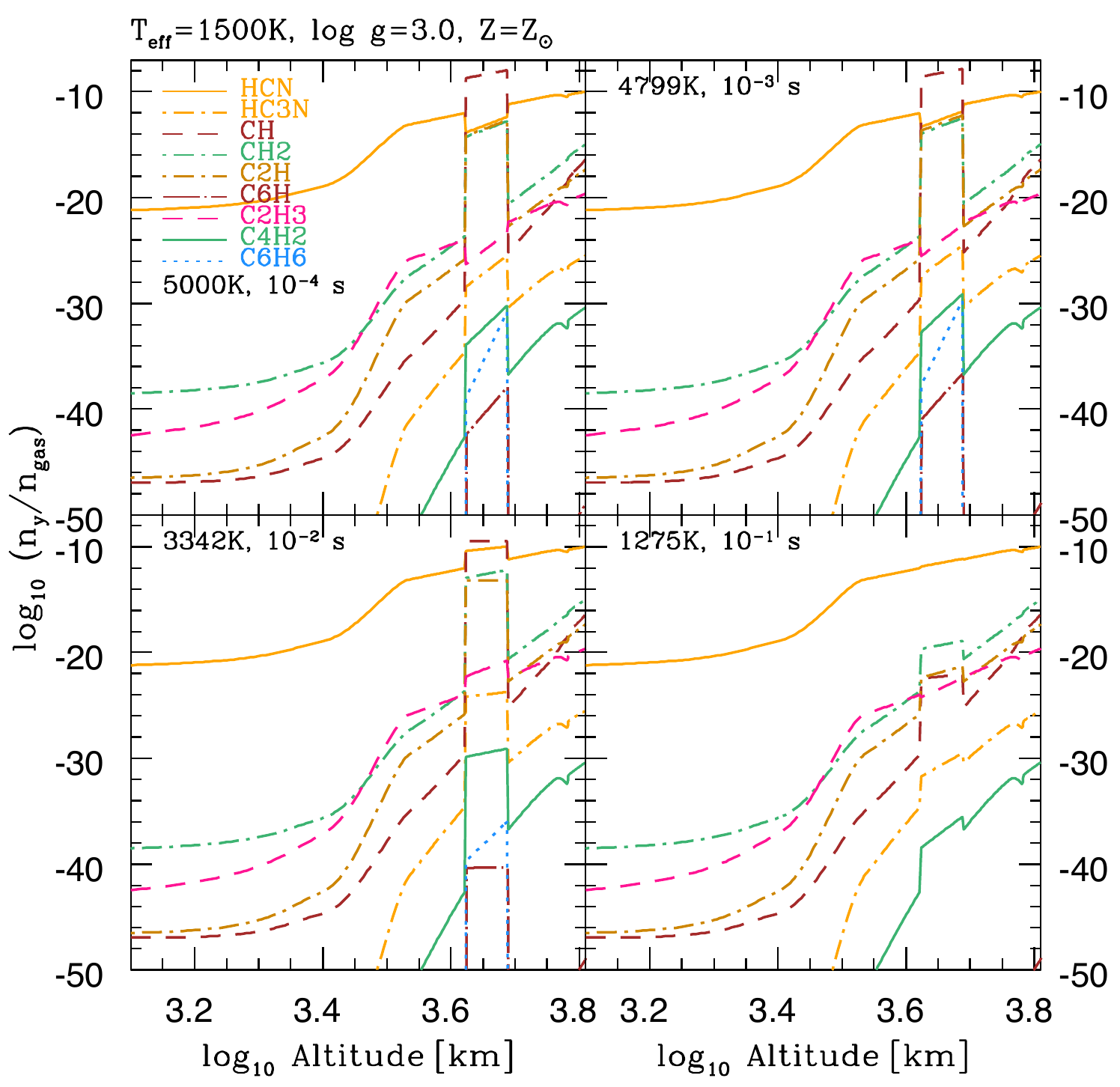}
\caption{Carbon molecules concentrations at different times after a
  lightning event in the cloud layer of a giant gas planet model
  atmosphere ($\rm T_{eff}=1500$, log(g)=3, solar). The figure panels
  follow the post-lightning temperature profile from Kov\'acs \&
  Tur\'anyi (2010).}
\label{fig:fourbox}
\end{figure}

\begin{figure}[htbp]
\vspace*{-0.5cm}
\includegraphics[scale=0.5]{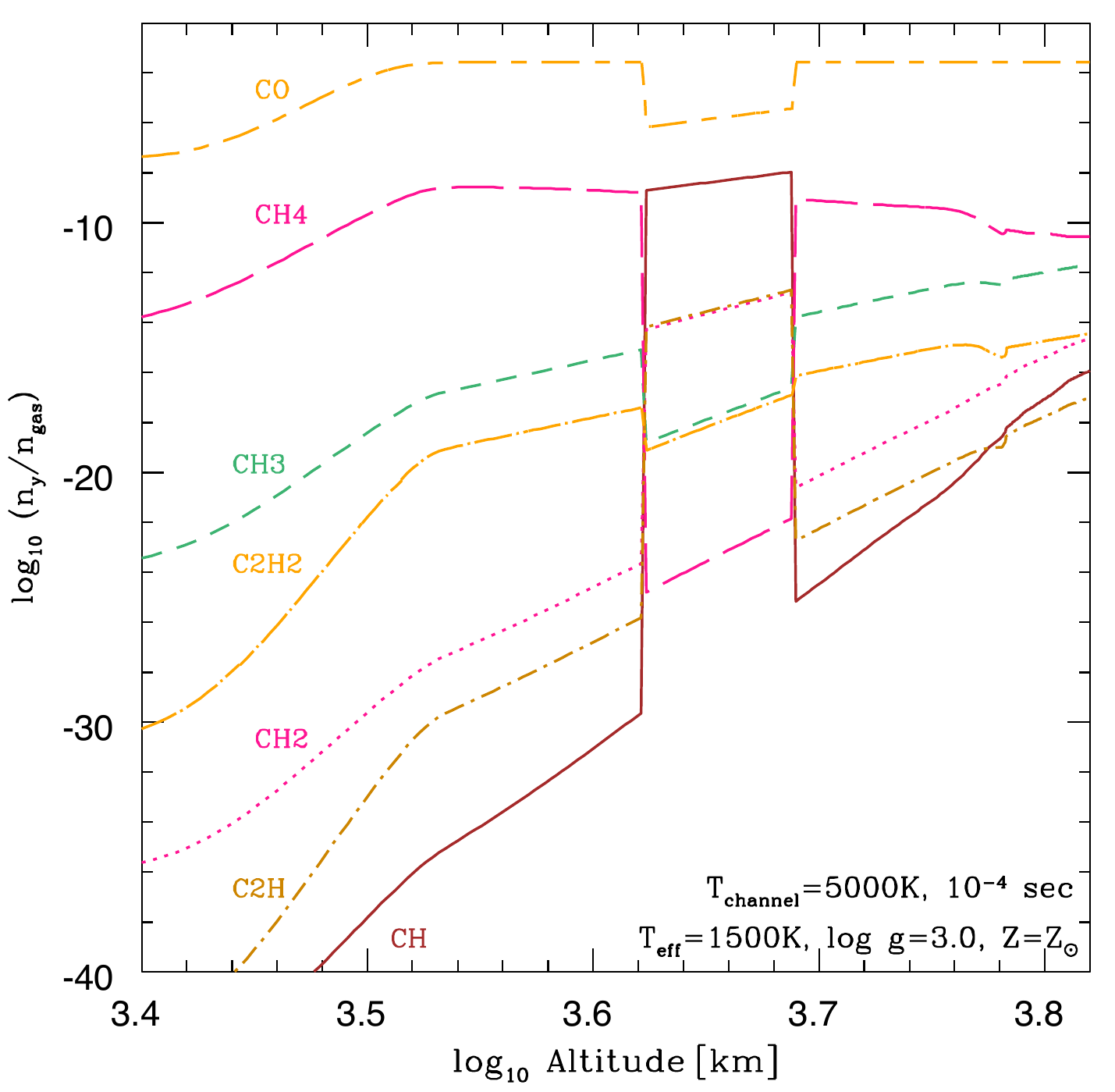}
\hspace*{0.5cm}
\includegraphics[scale=0.5]{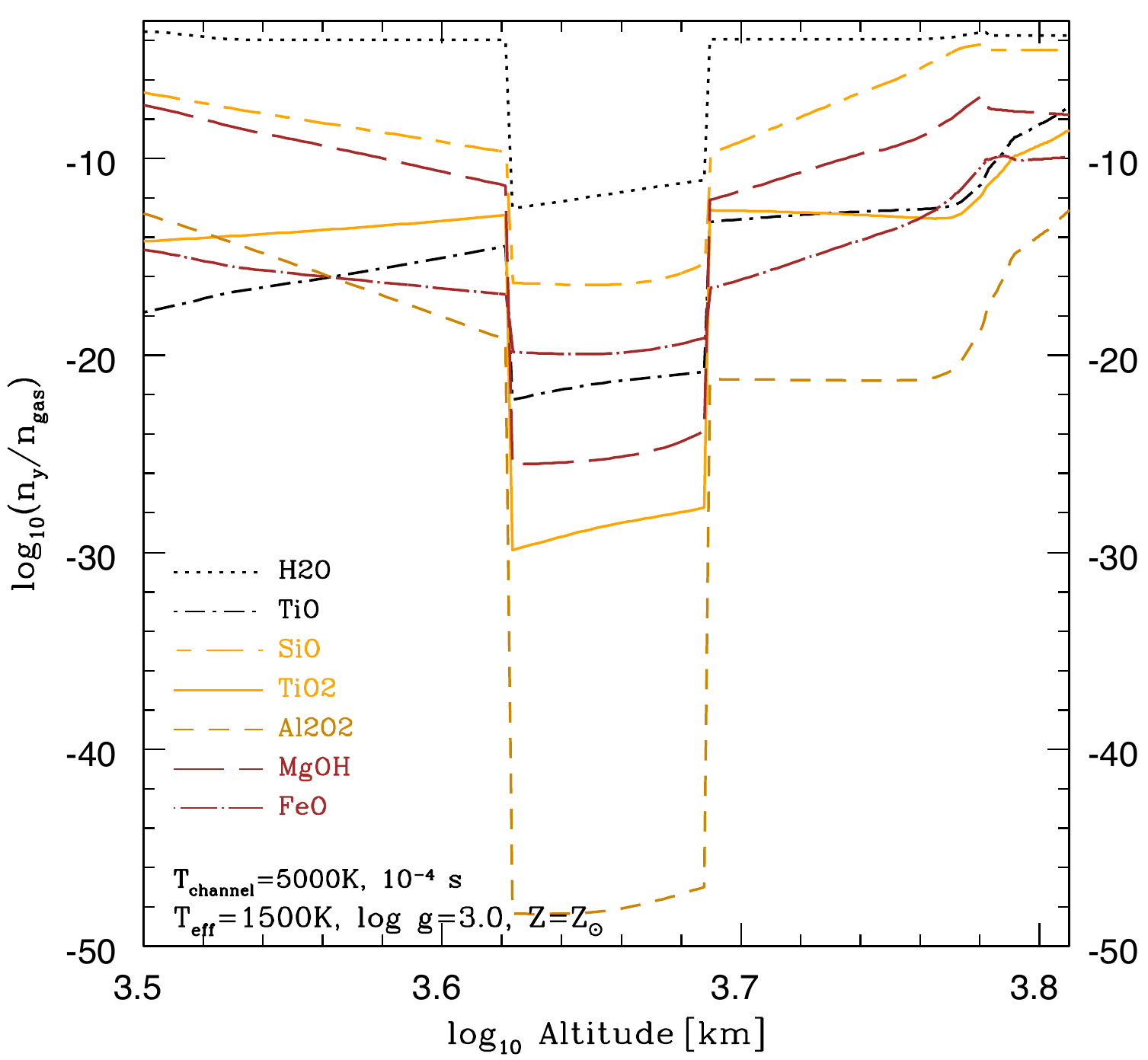}
\caption{{\bf Left:} A blow-up for the small hydrocarbon molecule concentration in a gas
 heated by  lightning discharge to $5000$K, $10^{-4}$ s after the
  initiation of the ionisation front. Note the increase of CH and C$_2$H in expense of CH$_4$.
{\bf Right:} Same for oxygen-binding molecules which participate in cloud particle condensation. H$_2$O is only indirectly affected by cloud formation, namely by the depletion of oxygen. }
\label{fig:light1}
\end{figure}

We started our chemistry calculations from a {\sc Drift-phoenix}
atmosphere model for a giant gas planet ($\rm T_{eff}=1500$ K,
log(g)=3, solar). We discuss possible changes of the gas-phase
composition due to changed thermal conditions of the atmospheric region
that was affected by a lighting discharge. The time-dependent
temperature decrease after the lightning event is directly adopted
from Fig. 1 in \citet{kovacs10}.

The lightning channel inside the dust cloud of our example gas-giant
atmosphere affected the atmospheric gas at heights bracketed by
T$_{\rm gas}=930$K and T$_{\rm gas}=1088$K (plotted as z-values in
Figs.~\ref{fig:fourbox} and~\ref{fig:light1}), which is equivalent to a
atmospheric volume of $\sim 10^{13}$ m$^3$.
We applied a chemical equilibrium routine to evaluate the vertical
profiles of C-bearing molecules (Figs. \ref{fig:fourbox},
\ref{fig:light1}) and of molecules that are involved in the cloud
particle condensation process (Fig.\ref{fig:light1}, right) by
mimicking the effect of an electrical breakdown by the enhancement of
the local gas temperature. Each panel in Figure \ref{fig:fourbox}
represents a different time after the electric discharge occurred, and
hence a different channel temperature: $5000$K ($0.0001$ seconds),
$4799$K ($0.001$ seconds), $3342$K ($0.01$ seconds) and $1275$K ($0.1$
seconds). The sharp edges correspond to the discontinuity in the
temperature profile, which suddenly rises to $5000$K, for example. The
production of certain carbonaceous species is hindered by these high
gas temperatures: the high temperature breaks the stable molecules
such as methane and carbon monoxide into radicals, which may recombine
to form molecules that are more stable at high temperatures (e.g. CH,
C$_2$H, HC$_3$N). These results generally agree with the lightning
model of \citet{kovacs10} for Titan. We did, however, not reproduce
the increase in the concentration of C$_2$H$_6$, and did not measure a
significant production of HCN, one of the proposed chemical products
of lightning (e.g. \citealt{lorenz08, hurley12}). This is an indication that the
chemical equilibrium approach is not appropriate for reactions
affecting these molecules.  Indeed, the after-lightning concentrations
of HCN, C$_2$H$_2$ and C$_2$H$_3$ only increase for a lower channel
temperature of 3342K.

The oxygen-bearing molecules underwent a decrease in their abundances
for every channel temperature considered, and also H$_2$O is
dissociated at such high temperatures. The decreased abundances of
molecules like TiO, SiO, SiO$_2$, MgOH, and FeO suggests a sudden stop
of potential condensation processes, hence, a freeze-in of the dust
properties (e.g. grain sizes), unless the high-temperature time span is
sufficiently long to evaporate the cloud particles completely.

The analysis presented here is a tentative investigation of the
effects of electrical discharges on atmospheric chemistry.  It is by
no means an exhaustive study. This work is a first step to see which
kind of differences in the gas-phase chemistry could be expected due
to the extreme temperature changes during a discharge event in
extrasolar atmospheres. Inspired by the Miller-Urey experiment, our
interest was to see if large carbon-binding macro-molecules could be
affected in the frame of our ansatz. This is a first step to identify
chemical species as a possible spectral fingerprint for extrasolar
lighting.

\section{On the observation of lightning in extrasolar atmospheres}\label{sec:propsobs}

Lightning and cosmic rays are important sources providing electrons to
the terrestrial atmosphere and the ionosphere. Lightning will
play a comparable role in Brown Dwarfs and exoplanets contributing to
the pool of non-thermal electrons that can accelerate along magnetic
field lines or that participate in coherent plasma motion, both
producing, for example, radio and X-ray emission. Such radio and X-ray
radiation was detected in brown dwarfs and is thought to origin from
accelerated electrons
(\citealt{berg2010,hall2002,burgasser2011,burgasser2013,route2013,williams2013,
  will2013b, cook2013}).

Nearby brown dwarfs that are not hiding inside a host-stars
magnetosphere are an easier target than exoplanets to detect radio
emission that originates from lightning directly (e.g. WISE1541:
2.8pc, GJ845Ba\&b: 3.6pc; \citealt{belu2013}). Further, the brown
dwarfs known to us are at much closer distances than known extrasolar
planets. The nearest brown dwarf is the binary (L8$\pm 1$ / T1$\pm 1$)
Luhman 16 (WISEJ1049) at a distance of $\sim$ 2pc which is a factor of
$5\times$ less than the distance assumed in
\citet{2012P&SS...74..156Z}. Luhman 16 (and other brown dwarfs) have
been suggested to show cloud-related variability \citep{gillon2013} in
the T-dwarf component, but no radio observations are available yet for
this object. The brown dwarf binary 2MASSJ1314 (L5/T7) has been
reported to show hyperactive quiescent radio emission
\citep{burgasser2013}. \cite{williams2013} present the first quiescent
radio emission from a T dwarf (T6.5, 2MASS J1047), which demonstrates
that radio emission is detectable from the coolest of the brown
dwarfs.  This radio emission is thought to arise from gyrosynchrotron
emission potentially pointing towards the presence of a weak
chromosphere, however, other mechanisms can not be rules out yet.
Non-equilibrium (like cosmic rays, \citealt{rimmer13}) or dynamic
ionisation processes (Alfv\'{e}n ionisation, \citealt{stark13}) were
suggested in addition to lightning discharges to play a significant
role in feeding the atmospheric {\it plasma} in ultra-cool, low-mass
objects to interact with its magnetosphere in order to explain the
origin of this radiation.

 To predict observable signature for lightning itself is not a trivial
 task as atmospheric electrical discharges produce broad-band observable
 signatures across the electromagnetic spectrum.
 Direct emission results from the afore mentioned acceleration of
 electrons (radio, X-ray and gamma-ray emission) and the excitation of
 atomic states, such as metastable states (IR, optical emission)
 during the discharge process. Indirect emission results from the
 effect on the local chemistry due to the electrified environment or
 the associated acoustic shock.  Additional emission as a consequence
 of secondary effects, such as the triggering of sprites and energetic
 electron beams, are also possible and have been detected for
 terrestrial lightning storms on Earth \citep{fuellekrug20013}.


\begin{singlespace}
\begin{table}[ht]
\tiny
\renewcommand{\arraystretch}{1.5}
\begin{threeparttable}
  \begin{tabular}{|l|l|l|l|l||l|}
	\hline
	Process & Signature & Wavelength & Celestial body & References & \vtop{\hbox{\strut Instrument with}\hbox{\strut suitable}\hbox{\strut wavelength range}} \\ 
	\hline \hline

	\vtop{\hbox{\strut Direct lightning}\hbox{\strut emission}} & \vtop{\hbox{\strut $\gamma$ - ray}\hbox{\strut (TGF)}} & \vtop{\hbox{\strut 20 eV - 40 MeV}} &  \multirow{6}{*}{Earth} & \vtop{\hbox{\strut \cite{lu2011, yair2012}}\hbox{\strut \cite{mar2010}}} & \vtop{\hbox{\strut Fermi GBM, \cite{mee2009}}\hbox{\strut AGILE, \cite{tav2006}}}\\ \cline{2-3} \cline{5-6}
	  
	  & X - ray & $30-250$ keV & & \vtop{\hbox{\strut \cite{dwy2004}}\hbox{\strut \cite{dwy2012}}}  &  \vtop{\hbox{\strut AGILE}\hbox{\strut Astrosat-SXT\tnote{1}}\hbox{\strut Astrosat-LAXPC\tnote{2}}} \\ \cline{2-6}
	 
	  & He & 588 nm & Jupiter & \vtop{\hbox{\strut Borucki et al. 1996}\hbox{\strut \cite{apl2013}}} & \vtop{\hbox{\strut VLT - X-SHOOTER}\hbox{\strut \cite{ver2011}}\hbox{\strut VLT - VIMOS,  \cite{lef2003}}} \\ \cline{2-6} 

	  & \vtop{\hbox{\strut NUV to NIR}\hbox{\strut many lines of}\hbox{\strut N$_2$, N(II),}\hbox{\strut O(I), O(II)}} & \vtop{\hbox{\strut See: \cite{wall1964}}\hbox{\strut (310-980 nm)}\hbox{\strut 0.35-0.85 $\mu$m}\hbox{\strut (direct imaging)}} & \vtop{\hbox{\strut Earth}\hbox{\strut Jupiter}} & \vtop{\hbox{\strut \cite{wall1964}}\hbox{\strut Baines et al. 2007}} & \vtop{\hbox{\strut Astrosat - UVIT, \cite{kum2012}}\hbox{\strut Swift-UVOT, \cite{rom2005}}\hbox{\strut VLT - X-SHOOTER}\hbox{\strut VLT - VIMOS}\hbox{\strut HARPS, \cite{may2003}}\hbox{\strut HST-NICMOS, \cite{via2009}}\hbox{\strut IRTF - TEXES, \cite{lac2002}}\hbox{\strut Spitzer IRS, \cite{hou2004}} } \\ \cline{2-6}

	  & whistlers & tens of Hz - kHz & \vtop{\hbox{\strut Earth}\hbox{\strut Saturn}\hbox{\strut Jupiter}} & \vtop{\hbox{\strut \cite{des2002}}\hbox{\strut \cite{yair2008, yair2012}}\hbox{\strut \cite{aka2006}}\hbox{\strut \cite{fis2008}}} & \vtop{\hbox{\strut LOFAR, \cite{vHar2013}}\hbox{\strut UTR 2, \cite{bar1978}}\hbox{\strut LWA, \cite{kass2005}}} \\ \cline{2-6}
	  
	  & sferics & 1 kHz - 100 MHz & \vtop{\hbox{\strut Earth}\hbox{\strut Saturn}\hbox{\strut Uranus}} & \vtop{\hbox{\strut \cite{des2002}}\hbox{\strut \cite{yair2008}}\hbox{\strut \cite{fis2008}}\hbox{\strut \cite{zarka86}}} & \vtop{\hbox{\strut LOFAR}\hbox{\strut UTR 2}\hbox{\strut LWA}} \\ 
	\hline \hline

	\vtop{\hbox{\strut Effect on}\hbox{\strut local}\hbox{\strut chemistry}} & NO$_x$ & \vtop{\hbox{\strut 439 nm (NO$_2$)}\hbox{\strut 445 nm (NO$_2$)}\hbox{\strut 5.3 $\mu$m (NO)}} & \vtop{\hbox{\strut Earth}\hbox{\strut Venus}} & \vtop{\hbox{\strut \cite{lorenz08}}\hbox{\strut \cite{nox1976}}\hbox{\strut \cite{kras2006}}} & \multirow{5}{*}{\vtop{\hbox{\strut HST-STIS}\hbox{\strut \cite{hern2012}}\hbox{\strut VLT -X-SHOOTER}\hbox{\strut VLT - VIMOS}\hbox{\strut HARPS}\hbox{\strut HST - NICMOS}\hbox{\strut IRTF-TEXES}\hbox{\strut Spitzer IRS}}} \\ \cline{2-5}
	  
 	  & O$_3$ & \vtop{\hbox{\strut $9.6$ $\mu$m}\hbox{\strut $14.3$ $\mu$m}\hbox{\strut $200-350$ nm}\hbox{\strut $420-830$ nm}\hbox{\strut }} & Earth & \vtop{\hbox{\strut \cite{tess2013}}\hbox{\strut }\hbox{\strut \cite{ehr2006}}} & \\ \cline{2-6}	  
	
	  & HCN & \vtop{\hbox{\strut 2.97525 $\mu$m}\hbox{\strut 3.00155 $\mu$m}} & \multirow{3}{*}{Jupiter} & \multirow{3}{*}{\vtop{\hbox{\strut\cite{des2002}}\hbox{\strut \cite{man2012}}}} & \multirow{3}{*}{\vtop{\hbox{\strut VLT - CRIRES, \cite{kau2004}}\hbox{\strut Keck - NIRSPEC, \cite{McL1998}}}} \\ \cline{2-3} 
	  
	  & C$_2$H$_2$ & \vtop{\hbox{\strut 2.998 $\mu$m}\hbox{\strut 3.0137 $\mu$m}} & & & \\
	\hline \hline
	  
	\multirow{4}{*}{\vtop{\hbox{\strut Emission caused}\hbox{\strut by secondary}\hbox{\strut events}\hbox{\strut (e.g. sprites)\tnote{3}}}} & 1PN$_2$ & $609-753$ nm & \multirow{5}{*}{Earth} & \multirow{5}{*}{\vtop{\hbox{\strut \cite{pasko07}}\hbox{\strut \cite{liu2007}}}} & \multirow{3}{*}{\vtop{\hbox{\strut HST - STIS}\hbox{\strut VLT - X-SHOOTER}\hbox{\strut VLT - VIMOS}\hbox{\strut HARPS}}} \\ \cline{2-3}
	  
	  & 1NN$_2^+$ & $391.4$ nm & & & \\ \cline{2-3}
	  
	  & 2PN$_2$ & $337$ nm & & & \\ \cline{2-3} \cline{6-6}
	  
	  & LBH N$_2$ & $150-280$ nm & & & \vtop{\hbox{\strut HST-COS,\cite{gre2012} }\hbox{\strut HST-STIS}} \\
	\hline
\end{tabular}
  \begin{tablenotes}
	\item[1] http://astrosat.iucaa.in/?q=node/14       
	\item[2] http://astrosat.iucaa.in/?q=node/12	
	\item[3] 1PN$_2$ is the first, 2PN$_2$ is the second positive,  LBH N$_2$ is the Lyman-Birge-Hopfield N$_2$ band system. 1NN$_2^+$ is the first negative band system of N$_2^+$. 
  \end{tablenotes}
\end{threeparttable}
\centering
\caption{Lightning discharges signatures observed in the Solar System. The right column lists potentially useful instruments  to observe lightning  
on extrasolar planets or brown dwarfs.}
\end{table}
\end{singlespace}

Research on discharge observables has focused on single events in the
Earth atmosphere and atmospheres in the solar system. Table
1\footnote{ Terrestrial Gamma-ray Flashes (TGFs) are brief (typically
  $<1$ ms) bursts of $\gamma$-rays with a mean energy of 2 MeV
  originating from the Earth's atmosphere (Lu et al. 2011). Both
  $\gamma$-ray and $X$-ray emissions are consequences of the
  production of energetic runaway electrons by lightning
  (\citealt{dwy2012}). {\it Sferics} (or atmospherics) are radio
  emissions in the low-frequency range with a power density peak at 10
  kHz on Earth.
Whistlers are electromagnetic waves propagating along magnetic field lines and emitting in the VLF range. 
The effect of the discharges on the local chemistry manifests in the higher abundance of certain  molecules.
}
summarises these signatures and links them to astronomical instruments according to their
wavelength capacity (right column). 
Which signature might appear in which exoplanet's or brown dwarf's atmosphere depends
on the atmospheric composition, temperature, element abundance, and
maybe velocity fields. 
The effect
of the discharges on the local chemistry manifests in the higher or
lower abundance of certain molecules\footnote{ For example, the
  largest natural source of nitrogen-oxyds in the Earth's troposphere
  is lightning (\citealt{yair2012}). \cite{kras2006} suggested that
  the observed NO abundance in Venus's atmosphere may be the result of
  lightning discharges.
On Earth, they typically emit in the blue and
red range of the spectrum.  \cite{bor1996} simulated the atmosphere of
a Jupiter-like planet and found that the He 588 nm line could appear
in the spectrum of a lightning discharge. \cite{tess2013} found that
ozone's 9.6 and 14.3 $\mu$m band could be significant in the atmosphere
of
super-Earths.} as we show in Sect.~\ref{s:chem}. A sensible assessment
of potential spectral signatures for planets other than solar-system-twins will require a radiative transfer
solution similar to the synthetic spectrum analysis done in model
atmosphere simulations.




\section{Concluding summery}\label{sec:concl}
Fossil evidence suggests lightning has influenced the Earth's atmosphere for at least 250
million years  \citep{harha66}. 
It has also been speculated that lightning could be responsible for
synthesising the first prebiotic molecules in a young Earth's
atmosphere \citep{miller53, miller59,johnson08}. \cite{stark13} demonstrate that prebiotic molecules could be synthesised on the surface of charged dust grains that are submerged in an atmospheric plasma. 
In order to assess
the potential role of lightning in brown dwarfs and extrasolar
planetary atmospheres, we applied laboratory and numerical
scaling laws for streamer discharges to atmospheres where mineral
clouds form. A comparison between the breakdown electric field and the
local electric field shows that electrical breakdown can occur at two
locations: inside the cloud layer (intra-cloud lightning) and above
the cloud layer (sprite). \citet{bocci95} observed that 80\% of the
sprites on Earth coincide with lightning ground strokes which was
confirmed by numerical modelling in combination with high-speed
measurements of sprite optical emissions (\citealt{liu09, gam11}).


From these locations in the atmosphere, the discharge propagates
through the atmosphere with subsequent branching until a minimum
diameter is reached and the discharge terminates. The total length of
the lightning strike generally extends over a longer distance in the
atmosphere of a brown dwarf than of a giant gas planet.  Consequently,
the atmospheric gas volume affected by one of these discharge events
(e.g.  by an increase of heat or an increase of the local number of
electrons) is larger in a brown dwarf atmosphere
($10^8~-~10^{10}$\,m$^3$) than in a giant gas planet's atmosphere
($10^4~-~10^{6}$\,m$^3$). The total dissipated energy in one event is
$<10^{12}$ J for all models of initial solar metallicity which is
below the values observed for Saturn and Jupiter. The dissipated
energy is higher in brown dwarf atmospheres than in giant gas
planetary atmospheres and increases with decreasing effective
temperature in both cases.

This all suggests that lightning events occurring in the atmospheres
of BDs may be easier to observe than on planets. However, the
likelihood of detection depends on the energetics of the discharge and
the proximity of its host to Earth. Many of the known exoplanets, in
particular those in the habitable zone, are considerably farther away
from Earth than the nearest brown dwarfs
(e.g. \citealt{LoCurto13}). Therefore, a non-detection is not evidence
for the absence of lightning on exoplanets.

The energy release by lightning is dissipated into the ambient medium
and, hence, causes an increase of the gas temperature inside the
discharge channel. Assuming that the chemistry remains in a steady
state (for all kinetic reactions $\tau_{\rm chem}<10^{-4}$s), we have
presented a tentative investigation of the impact lightning has on the
molecular composition of a GP atmospheric gas of solar metallicity. We
modelled the enhanced temperature in a localised atmospheric volume as
a result of the discharge event, and its effect on the local
chemistry.  First attempts to show the influence of lightning on the
local gas phase indicate an increase of small carbohydrate molecules
like CH and CH$_2$ at the expense of CO and CH$_4$. Dust forming
molecules are destroyed and the dust properties are frozen-in unless
enough time is available for complete evaporation.

The dissipated energy per lightning event calculated from our
simplified model is comparable to terrestrial values, and only reach
values comparable to Jupiter in a low-metallicity atmosphere which is
$\sim 10^4~-~10^5\times$ the terrestrial value.  As there is no
reason to doubt the fundamental physics of streamer propagation being
applicable  outside of our solar system, external factors like
rotation rates and cosmic rays may influence the occurrence of such
large-scale discharges. \citet{rimmer13} show that cosmic rays can
ionise the upper atmosphere of free-floating brown dwarfs and giant
gas planets, and cosmic rays may also trigger lightning
(\citealt{belog10,ryha2012,babich12}). The impact of cosmic rays will be
stronger for a brown dwarf because they are not shielded by a host star's
wind. Brown dwarfs can be rapid rotators (e.g. \citealt{scholz11})
which influences the wind speeds and the local conditions
for charge separation. Volcano plumes, which are composed of small
silicate ash particles, have a lightning activity that is order of
magnitudes larger than in a common thundercloud on Earth. These
arguments suggest that our results provide a lower limit for the
lightning dissipation energy and that lightning can be expected to be
stronger and more frequent in fast rotating extraterrestrial objects
that form mineral clouds.

{\bf Acknowledgement:} We thank the anonymous referee for a very
constructive refereeing process.  We thank Declan A. Diver,
M. F{\"u}llekrug, Aline Vidotto, Scott Gregory, and Peter Woitke for
helpful discussions. ChH, CRS, GH highlight financial support of the
European Community under the FP7 by an ERC starting grant. RB thanks
the Physics Trust of the University of St Andrews for supporting her
summer placement. Most literature search was performed using the
ADS. Our local computer support is highly acknowledged.

\bibliographystyle{apj}
\bibliography{bib}

\label{lastpage}



\end{document}